\documentclass[longauth]{aa}
\usepackage{amsmath}
\usepackage{color}
\usepackage{graphicx}
\usepackage{natbib}
\usepackage{txfonts}
\usepackage{url}
\usepackage{subfloat}
\usepackage[version=3]{mhchem}
\usepackage{CJK}
\usepackage{multirow}

\usepackage{natbib,twoopt}
\usepackage[breaklinks=true]{hyperref} 
\bibpunct{(}{)}{;}{a}{}{,}             
\hypersetup{
  colorlinks,
  citecolor=blue,
  linkcolor=blue,
  urlcolor=blue,
 }
\makeatletter
  \newcommandtwoopt{\citeads}[3][][]{\href{http://adsabs.harvard.edu/abs/#3}%
    {\def\hyper@linkstart##1##2{}%
     \let\hyper@linkend\@empty\citealp[#1][#2]{#3}}}
  \newcommandtwoopt{\citepads}[3][][]{\href{http://adsabs.harvard.edu/abs/#3}%
    {\def\hyper@linkstart##1##2{}%
     \let\hyper@linkend\@empty\citep[#1][#2]{#3}}}
  \newcommandtwoopt{\citetads}[3][][]{\href{http://adsabs.harvard.edu/abs/#3}%
    {\def\hyper@linkstart##1##2{}%
     \let\hyper@linkend\@empty\citet[#1][#2]{#3}}}
  \newcommandtwoopt{\citeyearads}[3][][]%
    {\href{http://adsabs.harvard.edu/abs/#3}
    {\def\hyper@linkstart##1##2{}%
     \let\hyper@linkend\@empty\citeyear[#1][#2]{#3}}}
\makeatother

\newcommand{\myemail}{\email{yangcht@pmo.ac.cn}}

\newcommand{\mum}{{\hbox {\,$\mu m$}}}

\newcommand{\hto}{{\hbox {H$_{2}$O}}}

\newcommand{\htop}{{\hbox {H$_2$O$^+$}}}

\newcommand{\ihto}{\hbox {$I_{\mathrm{H_2O}}$}}
\newcommand{\lhto}{\hbox {$L_{\mathrm{H_2O}}$}}
\newcommand{\lhtop}{\hbox {$L_{\mathrm{H_2O^+}}$}}
\newcommand{\nhto}{\hbox {$N_{\mathrm{H_2O}}$}}
\newcommand{\whto}{\hbox {$\Delta V_\mathrm{H_2O}$}}
\newcommand{\ico}{\hbox {$I_{\mathrm{CO}}$}}

\newcommand{\wco}{\hbox {$\Delta V_\mathrm{CO}$}}
\newcommand{\lir}{\hbox {$L_{\mathrm{IR}}$}}
\label{key}
\newcommand{\td}{{\hbox {$T_{\mathrm{d}}$}}}
\newcommand{\lsun}{{\hbox {$L_\odot$}}}

\def\t#1#2#3#4#5#6{{\hbox {$#1_{#2#3}\text{--}#4_{#5#6}$}}}
\def\tco#1#2{{\hbox {$#1\text{--}#2$}}}
\def\co#1#2{{\hbox {${\mathrm{CO}}(#1\text{--}#2)$}}}
\def\htot#1#2#3#4#5#6{\hbox {\hto(\t#1#2#3#4#5#6)}}
\def\lhtot#1#2#3#4#5#6{\hbox {$L_{\mathrm{H_2O}(#1_{#2#3}\text{--}#4_{#5#6})}$}}
\def\ihtot#1#2#3#4#5#6{\hbox {$I_{\mathrm{H_2O}(#1_{#2#3}\text{--}#4_{#5#6})}$}}
\def\lhtotlir#1#2#3#4#5#6{{\hbox {${L_{\mathrm{H_2O}(#1_{#2#3}\text{--}#4_{#5#6})}}\over{L_{\mathrm{IR}}}$}}}

\def\ihtotihto#1#2#3#4#5#6{{\hbox {${I_{\mathrm{H_2O}(3_{21}\text{--}3_{12})}}\over{I_{\mathrm{H_2O}\,(#1_{#2#3}\text{--}#4_{#5#6})}}$}}}

\defcitealias{2013ApJ...779...25B}{B13}
\defcitealias{2013A&A...551A.115O}{O13}
\defcitealias{2013ApJ...771L..24Y}{Y13}
\defcitealias{2014A&A...567A..91G}{G14}

\newcommand{\hz}{\hbox {high-redshift}}
\newcommand{\fir}{\hbox {far-infrared}}
\newcommand{\ir}{\hbox {infrared}}

\begin{document}
\begin{CJK*}{UTF8}{gbsn}

\title{Submillimeter \hto\ and \htop\ emission in lensed ultra- and hyper-luminous infrared 
       galaxies at $z\sim2$--4\thanks{{\it Herschel} is an ESA space observatory 
	   with science instruments provided by European-led Principal Investigator 
	   consortia and with important participation from NASA.}
	   }
\author
{
 C. Yang (杨辰涛)\inst{1,2,3,4,5} \and 
 A. Omont\inst{4,5}              \and 
 A. Beelen\inst{2}               \and 
 E. Gonz{\'a}lez-Alfonso\inst{6} \and 
 R. Neri\inst{7}                 \and 
 Y. Gao (高煜)\inst{1}            \and 
 P. van der Werf\inst{8}         \and
 A. Wei{\ss}\inst{9}             \and
 R. Gavazzi\inst{4,5}            \and
 N. Falstad\inst{10}             \and 
 A. J. Baker\inst{11}            \and
 R. S. Bussmann\inst{12}         \and
 A. Cooray\inst{13}              \and 
 P. Cox\inst{14}                 \and
 H. Dannerbauer\inst{15}         \and \\
 S. Dye\inst{16}                 \and
 M. Gu{\'e}lin\inst{7}           \and
 R. Ivison\inst{17,18}           \and
 M. Krips\inst{7}                \and
 M. Lehnert\inst{4,5}            \and
 M.J. Micha\l{}owski\inst{17}    \and \\
 D.A. Riechers\inst{12}          \and
 M. Spaans\inst{19}              \and
 E. Valiante\inst{20}
}

\institute{
Purple Mountain Observatory/Key Lab of Radio Astronomy, Chinese Academy of Sciences, 
Nanjing 210008, PR China \\
\myemail
\and
Institut d$'$Astrophysique Spatiale, CNRS, Univ. Paris-Sud, Universit\'{e} Paris-Saclay, 
B\^{a}t. 121, 91405 Orsay cedex, France
\and
Graduate University of the Chinese Academy of Sciences, 19A Yuquan Road, 
Shijingshan District, 10049, Beijing, PR China
\and
CNRS, UMR 7095, Institut d$'$Astrophysique de Paris, F-75014, Paris, France 
\and  
UPMC Univ. Paris 06, UMR 7095, Institut d$'$Astrophysique de Paris, 
F-75014, Paris, France
\and  
Universidad de Alcal{\'a}, Departamento de Física y Matem{\'a}ticas, Campus 
Universitario, 28871 Alcal{\'a} de Henares, Madrid, Spain
\and
Institut de Radioastronomie Millim{\'e}trique (IRAM), 300 rue de la Piscine, 
38406 Saint-Martin-d’H{\`e}res, France  
\and
Leiden Observatory, Leiden University, P.O. Box 9513, NL-2300 RA Leiden, 
The Netherlands
\and
Max Planck Institut f{\"u}r Radioastronomie, Auf dem Hgel 69,
53121 Bonn
\and
Department of Earth and Space Sciences, Chalmers University of Technology, 
Onsala Space Observatory, 43992 Onsala, Sweden
\and
Department of Physics and Astronomy, Rutgers, The State University of New Jersey, 
136 Frelinghuysen Road, Piscataway, NJ 08854-8019, USA
\and
Astronomy Department, Cornell University, 220 Space Sciences Building, 
Ithaca, NY 14853, USA
\and
Department of Physics and Astronomy, University of California, 
Irvine, Irvine, CA 92697
\and
Joint ALMA Observatory, Alonso de C{\'o}rdova 3107, Vitacura, Santiago, Chile
\and
Universitat Wien, Institut f{\"u}r Astrophysik, T{\"u}rkenschanzstrasse 17, 
1180 Wien, Austria
\and
School of Physics and Astronomy, University of Nottingham, University Park, 
Nottingham NG7 2RD, UK
\and
Institute for Astronomy, University of Edinburgh, Royal Observatory, 
Blackford Hill, Edinburgh, EH9 3HJ, UK
\and
European Southern Observatory, Karl Schwarzschild Stra{\ss}e 2, 85748, Garching, Germany
\and
Kapteyn Astronomical Institute, University of Groningen, 
9700 AV Groningen, The Netherlands
\and
School of Physics and Astronomy, Cardiff University, The Parade, 
Cardiff CF24 3AA, UK
}

\date {Received .../ Accepted ...}

\abstract
{
We report rest-frame submillimeter \hto\ emission line observations of 11 ultra- 
or hyper-luminous infrared galaxies (ULIRGs or HyLIRGs) at $z\,\sim\,2\text{--}4$ 
selected among the brightest lensed galaxies discovered in the {\it Herschel}-Astrophysical 
Terahertz Large Area Survey ({\it H}-ATLAS). Using the IRAM NOrthern Extended 
Millimeter Array (NOEMA), we have detected 14 new \hto\ emission lines. These 
include five \t321312 ortho-\hto\ lines ($E_\mathrm{up}$/$k = 305\,$K) and nine 
$J=2$ para-\hto\ lines, either \t202111 ($E_\mathrm{up}$/$k = 101\,$K) or \t211202 
($E_\mathrm{up}$/$k=137\,$K). The apparent luminosities of the \hto\ emission lines 
are $\mu$\lhto$\,\sim 6\text{--}21 \times 10^8$\,\lsun\, ($3<\mu<15$, where $\mu$ 
is the lens magnification factor), with velocity-integrated line fluxes ranging 
from $4\text{--}15$ Jy\,km\,s$^{-1}$. We have also observed CO emission lines using 
EMIR on the IRAM 30m telescope in seven sources (most of those have not yet had 
their CO emission lines observed). The velocity widths for CO and \hto\ lines 
are found to be similar, generally within 1\,$\sigma$ errors in the same source.  
With almost comparable integrated flux densities to those of the high-$J$ CO 
line (ratios range from 0.4 to 1.1), \hto\ is found to be among the strongest 
molecular emitters in \hz\ Hy/ULIRGs. We also confirm our previously found 
correlation between luminosity of \hto\ (\lhto) and infrared (\lir) that 
\lhto\,$\sim$\lir$^{1.1\text{--}1.2}$, with our new detections. This correlation 
could be explained by a dominant role of far-infrared pumping in the \hto\ 
excitation. Modelling reveals that the far-infrared radiation fields have 
warm dust temperature $T_\mathrm{warm} \sim 45\text{--}75$\,K, \hto\ column 
density per unit velocity interval 
\nhto/$\Delta V \gtrsim 0.3\times10^{15}$\,cm$^{-2}$\,km$^{-1}$\,s and 100\,$\mu$m 
continuum opacity $\tau_{100} > 1$ (optically thick), indicating that \hto\ is likely 
to trace highly obscured warm dense gas. However, further observations of $J\geq4$ 
\hto\ lines are needed to better constrain the continuum optical depth and other 
physical conditions of the molecular gas and dust. We have also detected \htop\ 
emission in three sources. A tight correlation between \lhto\ and \lhtop\ has 
been found in galaxies from low to high redshift. The velocity-integrated flux 
density ratio between \htop\ and \hto\ suggests that cosmic rays generated by 
strong star formation are possibly driving the \htop\ formation.
}


\authorrunning{C. Yang et al.}

\titlerunning{\hto\ excitation in lensed Hy/ULIRGs at $z \sim 2\text{--}4$}

\maketitle

\section{Introduction}
\label{section:Introduction}
After molecular hydrogen (H$_2$) and carbon monoxide (CO), the water molecule (H$_2$O) 
can be one of the most abundant molecules in the interstellar medium (ISM) in galaxies. 
It provides some important diagnostic tools for various physical and chemical 
processes in the ISM \citep[e.g.][and references therein]{2013ChRv..113.9043V}. 
Prior to the {\it Herschel Space Observatory} \citep{2010A&A...518L...1P}, in 
extragalactic sources, non-maser \hto\ rotational transitions were only detected by the 
\textit{Infrared Space Observatory} \citep[\textit{ISO},][]{1996A&A...315L..27K} in the 
form of far-infrared absorption lines \citep{2004ApJ...613..247G, 2008ApJ...675..303G}. 
Observations of local infrared bright galaxies by {\it Herschel} have revealed a rich 
spectrum of submillimeter (submm) \hto\ emission lines (submm \hto\ refers to 
rest-frame submillimeter \hto\ emission throughout this paper if not otherwise specified). 
Many of these lines are emitted from high-excitation rotational levels with upper-level 
energies up to $E_\mathrm{up}$/$k = 642\,$K \citep[e.g.][]{2010A&A...518L..42V, 
2010A&A...518L..43G, 2012A&A...541A...4G, 2013A&A...550A..25G, 2011ApJ...743...94R, 
2012ApJ...753...70K, 2012ApJ...758..108S, 2013ApJ...762L..16M, 2013ApJ...779L..19P, 
2013ApJ...768...55P}. Excitation analysis of these lines has revealed that they are 
probably excited through absorption of \fir\ photons from thermal dust emission in 
warm dense regions of the ISM \citep[e.g.][]{2010A&A...518L..43G}. Therefore, 
unlike the canonical CO lines that trace collisional excitation of the molecular 
gas, these \hto\ lines represent a powerful diagnostic of the \fir\ radiation field.

\begin{figure}[htbp]
	 \begin{center}
 \includegraphics[scale=0.441]{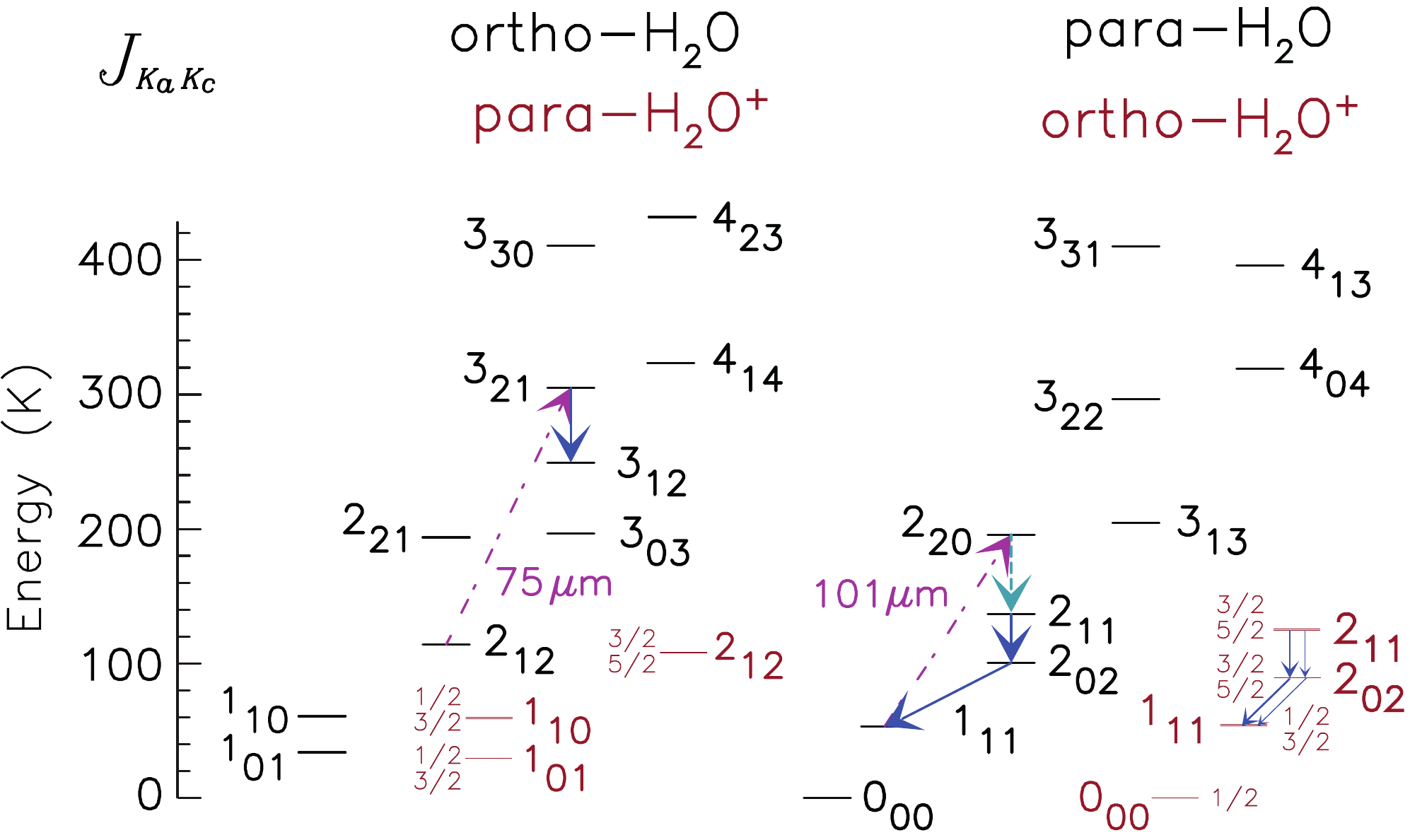}
 \caption{
          Energy level diagrams of \hto\ and \htop\ shown in black and red, 
          respectively. Dark blue arrows are the submm \hto\ transitions we 
		  have observed in this work. Pink dashed lines show the \fir\ pumping 
		  path of the \hto\ excitation in the model we use, with the wavelength 
		  of the photon labeled. The light blue dashed arrow is the transition 
		  from para-\hto\ energy level $2_{20}$ to $2_{11}$ along the cascade 
		  path from $2_{20}$ to $1_{11}$. Rotational energy levels of \hto\ 
		  and \htop, as well as fine structure component levels of \htop\, are 
		  also shown in the figure.
         }
 \label{fig:h2o-e-level}
     \end{center}
 \end{figure}

Using the {\it Herschel} archive data, \citet[][hereafter Y13]{2013ApJ...771L..24Y} 
have undertaken a first systematic study of submm \hto\ emission in local \ir\ 
galaxies. \hto\ was found to be the strongest molecular emitter after CO within 
the submm band in those \ir-bright galaxies, even with higher flux density than 
that of CO in some local ULIRGs (velocity-integrated flux density of \htot321312 is 
larger than that of \co54 in four galaxies out of 45 in the \citetalias{2013ApJ...771L..24Y} 
sample). The luminosities of the submm \hto\ lines (\lhto) are near-linearly 
correlated with total \ir\ luminosity ($L_\mathrm{IR}$, integrated over 
8--1000\,$\mu$m) over three orders of magnitude. The correlation is revealed 
to be a straightforward result of \fir\ pumping: \hto\ molecules are excited 
to higher energy levels through absorbing \fir\ photons, then the upper level 
molecules cascade toward the lines we observed in an almost constant fraction 
(Fig.\,\ref{fig:h2o-e-level}). Although the galaxies dominated by active galactic 
nuclei (AGN) have somewhat lower ratios of \lhto/\lir, there does not appear 
to be a link between the presence of an AGN and the submm \hto\ emission 
\citepalias{2013ApJ...771L..24Y}. The \hto\ emission is likely to trace the 
\fir\ radiation field generated in star-forming nuclear regions in galaxies, 
explaining its tight correlation with \fir\ luminosity.

Besides detections of the \hto\ lines in local galaxies from space telescopes, 
redshifted submm \hto\ lines in \hz\ lensed Ultra- and Hyper-Luminous InfraRed 
Galaxies (ULIRGs, $10^{13}\,L_\odot > L_\mathrm{IR} \geq 10^{12}$\,\lsun; HyLIRGs, 
$L_\mathrm{IR} \geq 10^{13}$\,\lsun) can also be detected by ground-based 
telescopes in atmospheric windows with high transmission. Strong gravitational 
lensing boosts the flux and allows one to detect the \hto\ emission lines easily. 
Since our first detection of submm \hto\ in a lensed {\it Herschel} source at 
$z = 2.3$ \citep{2011A&A...530L...3O} using the IRAM NOrthern Extended Millimeter 
Array (NOEMA), several individual detections at \hz\ have also been reported 
\citep{2011ApJ...738L...6L, 2011ApJ...741L..38V, 2011ApJ...741L..37B, 
2012A&A...538L...4C, 2012ApJ...757..135L, 2013ApJ...779...67B, 
2013A&A...551A.115O, 2013Natur.495..344V, 2013ApJ...767...88W,
2014ApJ...783...59R}. These numerous and easy detections of \hto\ in \hz\ 
lensed ULIRGs show that its lines are the strongest submm molecular lines 
after CO and may be an important tool for studying these galaxies.

We have carried out a series of studies focussing on submm 
\hto\ emission in \hz\ lensed galaxies since our first detection. Through the 
detection of $J=2$ \hto\ lines in seven \hz\ lensed Hy/ULIRGs reported by 
\citet[][hereafter O13]{2013A&A...551A.115O}, a slightly super-linear correlation 
between \lhto\ and \lir\ (\lhto\;$\propto$\;\lir$^{1.2}$) from local ULIRGs and 
\hz\ lensed Hy/ULIRGs has been found. This result may imply again that \fir\ 
pumping is important for \hto\ excitation in \hz\ extreme starbursts. The 
average ratios of \lhto\ to \lir\ for the $J=2$ \hto\ lines in the \hz\ sources 
tend to be $1.8\pm0.9$ times higher 
than those seen locally \citepalias{2013ApJ...771L..24Y}. This shows that the 
same physics with infrared pumping should dominate \hto\ excitation in ULIRGs 
at low and high redshift, with some specificity at \hz\ probably linked to 
the higher luminosities.

Modelling provides additional information about the \hto\ excitation. For example,
through LVG modelling, \cite{2013Natur.496..329R} argue that the excitation of the 
submm \hto\ emission in the $z \sim 6.3$ submm galaxy is \fir\ pumping dominated. 
Modelling of the local {\it Herschel} galaxies of \citetalias{2013ApJ...771L..24Y} 
has been carried out by \citet[][hereafter G14]{2014A&A...567A..91G}. They confirm 
that \fir\ pumping is the dominant mechanism responsible for the submm \hto\ 
emission (except for the ground-state emission transitions, such as para-\hto\ 
transition \t111000) in the extragalactic sources. Moreover, collisional 
excitation of the low-lying ($J \leq 2$) \hto\ 
lines could also enhance the radiative pumping of the ($J \geq 3$) high-lying 
lines. The ratio between low-lying and high-lying \hto\ lines is sensitive to the 
dust temperature (\td) and \hto\ column density ($N_\mathrm{H_2O}$). From modelling 
the average of local star-forming- and mild-AGN-dominated galaxies, 
\citetalias{2014A&A...567A..91G} show that the submm \hto\ emission comes from 
regions with $N_\mathrm{H_2O} \sim (0.5\text{--}2) \times 10^{17}$\,cm$^{-2}$ and a 
100\,$\mu$m continuum opacity of $\tau_{100} \sim 0.05\text{--}0.2$, where 
\hto\ is mainly excited by warm dust with a temperature range of $45\text{--}75$\,K.
\hto\ lines thus provide key information about the properties of the dense 
cores of ULIRGs, that is, their \hto\ content, the infrared radiation field and the 
corresponding temperature of dust that is warmer than the core outer layers 
and dominates the far-infrared emission.

Observations of the submm \hto\ emission, together with appropriate modelling and 
analysis, therefore allows us to study the properties of the \fir\ radiation 
sources in great detail. So far, the excitation analysis combining both 
low- and high-lying \hto\ emission has only been done in a few case studies. 
Using \hto\ excitation modelling considering both collision and \fir\ pumping, 
\cite{2010A&A...518L..43G} and \cite{2011ApJ...741L..38V} estimate the sizes of 
the \fir\ radiation fields in Mrk\,231 and APM\,08279+5255 (APM\,08279 hereafter), 
which are not resolved by the observations directly, and suggest their AGN dominance 
based on their total enclosed energies. This again demonstrates that submm \hto\ 
emission is a powerful diagnostic tool which can even transcend the angular resolution 
of the telescopes.

The detection of submm \hto\ emission in the {\it Herschel}-ATLAS\footnote{The 
{\it Herschel}-ATLAS is a project with {\it Herschel}, which is an ESA space 
observatory with science instruments provided by European-led Principal Investigator 
consortia and with important participation from NASA. The {\it H}-ATLAS website is 
\url{http://www.h-atlas.org}.} \citep[][{\it H}-ATLAS hereafter]{2010PASP..122..499E} 
sources through gravitational lensing allows us to characterise the \fir\ radiation 
field generated by intense star-forming activity, and possibly AGN, and learn the 
physical conditions in the warm dense gas phase in extreme starbursts in the early 
Universe. Unlike standard dense gas tracers such as HCN, which is weaker at \hz\ 
compared to that of local ULIRGs \citep{2007ApJ...660L..93G}, submm \hto\ lines are 
strong and even comparable to high-$J$ CO lines in some galaxies 
\citepalias{2013ApJ...771L..24Y, 2013A&A...551A.115O}. Therefore, \hto\ is an 
efficient tracer of the warm dense gas phase that makes up a major fraction of the 
total molecular gas mass in \hz\ Hy/ULIRGs \citep{2014PhR...541...45C}. The successful 
detections of submm \hto\ lines in both local \citepalias{2013ApJ...771L..24Y} and the 
\hz\ universe \citepalias{2013A&A...551A.115O} show the great potential of a systematic 
study of \hto\ emission in a large sample of \ir\ galaxies over a wide range in redshift 
(from local up to $z\sim4$) and luminosity ($\lir \sim10^{10}$--$10^{13}$\,\lsun). However,
our previous \hz\ sample was limited to seven sources and to one $J=2$ para-\hto\ line 
($E_\mathrm{up}$/$k = 100$--$127\,$K) per source \citepalias{2013A&A...551A.115O}. In 
order to further constrain the conditions of \hto\ excitation, to confirm the dominant 
role of \fir\ pumping and to learn the physical conditions of the warm dense gas phase 
in \hz\ starbursts, it is essential to extend the studies to higher excitation lines. 
We thus present and discuss here the results of such new observations of a strong $J=3$ 
ortho-\hto\ line with $E_\mathrm{up}$/$k = 304\,$K in six strongly lensed {\it H}-ATLAS 
galaxies at z\,$\sim$\,2.8--3.6, where a second lower-excitation $J=2$ para-\hto\ line 
was also observed (Fig.\,\ref{fig:h2o-e-level} for the transitions and the 
corresponding $E_\mathrm{up}$).

We describe our sample, observation and data reduction in Section \ref{section:Sample 
and Observation}. The observed properties of the \hz\ submm \hto\ emission are 
presented in Section \ref{Results}. Discussions of the lensing properties, \lhto-\lir\ 
correlation, \hto\ excitation, comparison between \hto\ and CO, AGN contamination will 
be given in Section \ref{Discussion}. Section \ref{htop} describes the detection of 
\htop\ lines. We summarise our results in Section \ref{Conclusions}. A flat $\Lambda$CDM 
cosmology with 
$H_{0}=71\,{\rm km\,s^{-1}\,Mpc^{-1}}$, $\Omega_{M}=0.27$, $\Omega_{\Lambda}=0.73$ 
\citep{2003ApJS..148..175S} is adopted throughout this paper.

\section{Sample and observation}
\label{section:Sample and Observation}
Our sample consists of eleven extremely bright \hz\ sources with 
$F_\mathrm{500\mu m}>200$\,mJy discovered by the {\it H}-ATLAS 
survey \citep{2010PASP..122..499E}. Together with the seven similar 
sources reported in our previous H$_2$O study \citepalias{2013A&A...551A.115O}, 
they include all the brightest \hz\ {\it H}-ATLAS sources ($F_\mathrm{500\mu m}>170$\,mJy), 
but two, imaged at 880\,$\mu$m with SMA by \citet[][hereafter B13]{2013ApJ...779...25B}. 
In agreement with the selection according to the methods of 
\citet{2010Sci...330..800N}, the detailed lensing modelling 
performed by \citetalias{2013ApJ...779...25B} has shown that 
all of them are strongly lensed, but one, G09v1.124 \citep[][see below]{2013ApJ...772..137I}. 
The sample of our present study is thus well representative 
of the brightest \hz\ submillimeter sources with $F_\mathrm{500\mu m}>200$\,mJy 
(with apparent total infrared luminosity $\sim 5\text{--}15 \times 10^{13}$\,\lsun\ 
and $z \sim 1.5\text{--}4.2$) found by {\it H}-ATLAS in its equatorial ('GAMA') 
and north-galactic-pole ('NGP') fields, in $\sim 300$\,deg$^2$ with a density 
$\sim 0.05$\,deg$^{-2}$. In our previous project \citepalias{2013A&A...551A.115O}, 
we observed \hto\ in seven strongly lensed \hz\ {\it H}-ATLAS galaxies from the 
\citetalias{2013ApJ...779...25B} sample. In this work, in order to observe 
the high-excitation ortho-\htot321312 line with rest frequency of 1162.912\,GHz 
with the IRAM/NOEMA, we selected the brightest sources at 500\,$\mu$m with 
$z \gtrsim 2.8$ so that the redshifted lines could be observed in a reasonably 
good atmospheric window at $\nu_\mathrm{obs} \lesssim 300$\,GHz. Eight sources 
with such redshift were selected from the \citetalias{2013ApJ...779...25B} 
{\it H}-ATLAS sample.

\begin{table*}[htbp]
\small
\setlength{\tabcolsep}{0.42em}
\caption{Observation log.}
\centering
\begin{tabular}{lcllllllll}
\hline
\hline
{IAU Name}                                        & {Source}      & {$\mathrm{RA}$}              & {$\mathrm{DEC}$}                &  {$\mathrm{RA_{pk}}$} &  {$\mathrm{DEC_{pk}}$} &  {\hto\ line}             &{$\nu_\mathrm{obs}$}     & {Beam}                         &{$t_\mathrm{on}$} \\
                                                  &               &  (J2000)                     & (J2000)                         &   (J2000)             &  (J2000)               &                           &  (GHz)                  & ($''$)                         &  (h)                 \\
\hline                                                                                                                                                                                                                                                                                                                     
{{\it H}-ATLAS J083051.0$+$013224}                & {G09v1.97}    & {08:30:51.02}                & {$+$01:32:24.88}                &  {08:30:51.17}        &  {$+$01:32:24.39}      &  \t211202                 &  {162.286}              & {5.6$\times$3.3}               &  {3.5}               \\
                                                  &               &                              &                                 &  {08:30:51.17}        &  {$+$01:32:24.09}      &  \t321312                 &  {250.952}              & {2.6$\times$1.1}               &  {3.1}               \\
                                                                                                                                                                                                                                                                                                
{{\it H}-ATLAS J113526.3$-$014605}                & {G12v2.43}    & {11:35:26.36}                & {$-$01:46:05.56}                &  {11:35:26.27}        &  {$-$01:46:06.44}      &  \t202111                 &  {239.350}              & {2.3$\times$1.0}               &  {6.9}               \\
                                                  &               &                              &                                 &  {11:35:26.28}        &  {$-$01:46:06.43}      &  \t321312                 &  {281.754}              & {2.2$\times$1.1}               &  {1.5}               \\
                                                                                                                                                                                                                                                                                                
{{\it H}-ATLAS J125632.7$+$233625}                & {NCv1.143}    & {12:56:32.70}                & {$+$23:36:24.86}                &  {12:56:32.56}        &  {$+$23:36:27.92}      &  \t211202                 &  {164.739}              & {3.1$\times$2.9}               &  {1.5}               \\
                                                  &               &                              &                                 &  {12:56:32.56}        &  {$+$23:36:27.69}      &  \t321312                 &  {254.745}              & {2.1$\times$1.0}               &  {1.5}               \\
                                                                                                                                                                                                                                                                                                
{{\it H}-ATLAS J132630.1$+$334410}                & {NAv1.195}    & {13:26:30.12}                & {$+$33:44:09.90}                &  {13:26:30.14}        &  {$+$33:44:09.11}      &  \t202111                 &  {250.045}              & {2.0$\times$1.7}               &  {3.8}               \\
                                                  &               &                              &                                 &  {13:26:30.14}        &  {$+$33:44:09.09}      &  \t321312                 &  {293.334}              & {1.0$\times$0.9}               &  {3.1}               \\
                                                                                                                                                                                                                                                                                                
{{\it H}-ATLAS J132859.3$+$292327}                & {NAv1.177}    & {13:28:59.29}                & {$+$29:23:27.07}                &  {13:28:59.25}        &  {$+$29:23:26.18}      &  \t202111                 &  {261.495}              & {1.9$\times$1.7}               &  {2.3}               \\
                                                  &               &                              &                                 &  {13:28:59.25}        &  {$+$29:23:26.34}      &  \t321312                 &  {307.812}              & {1.6$\times$0.9}               &  {2.3}               \\
                                                                                                                                                                                                                                                                                                
{{\it H}-ATLAS J133008.4$+$245900}                & {NBv1.78}     & {13:30:08.56}                & {$+$24:58:58.30}                &  {13:30:08.56}        &  {$+$24:58:58.55}      &  \t321312                 &  {282.878}              & {1.7$\times$1.1}               &  {4.2}               \\                                                                                                                                
\multirow{2}{*}{{\it H}-ATLAS J084933.4$+$021443} & {G09v1.124-W} & \multirow{2}{*}{08:49:33.36} & \multirow{2}{*}{$+$02:14:42.30} &  {08:49:33.59}       &  {$+$02:14:44.68}     & \multirow{2}{*}{\t211202} &\multirow{2}{*}{220.537} &\multirow{2}{*}{1.8$\times$1.2} &\multirow{2}{*}{8.4}  \\
                                                  & {G09v1.124-T} &                              &                                 &  {08:49:32.95}       &  {$+$02:14:39.70}     &                           &                         &                                &                      \\
{{\it H}-ATLAS J085358.9$+$015537}                & {G09v1.40}    & {08:53:58.90}                & {$+$01:55:37.00}                &  {08:53:58.84}        &  {$+$01:55:37.75}      &  \t211202                 &  {243.425}              & {1.8$\times$1.0}               &  {1.9}               \\
{{\it H}-ATLAS J091043.1$-$000321}                & {SDP11}       & {09:10:43.09}                & {$-$00:03:22.51}                &  {09:10:43.06}        &  {$-$00:03:22.10}      &  \t202111                 &  {354.860}              & {1.9$\times$1.5}               &  {3.8}               \\
{{\it H}-ATLAS J125135.4$+$261457}                & {NCv1.268}    & {12:51:35.46}                & {$+$26:14:57.52}                &  {12:51:35.38}        &  {$+$26:14:58.12}      &  \t211202                 &  {160.864}              & {2.9$\times$2.6}               &  {7.7}               \\
{{\it H}-ATLAS J134429.4$+$303036}                & {NAv1.56}     & {13:44:29.52}                & {$+$30:30:34.05}                &  {13:44:29.46}        &  {$+$30:30:34.01}      &  \t211202                 &  {227.828}              & {1.7$\times$1.7}               &  {2.3}               \\
\hline
\end{tabular}
\tablefoot{
        RA and DEC are the J2000 {\it Herschel} coordinates which were taken as the centres of 
        the NOEMA images displayed in Fig.\,\ref{fig:map-all}; RA$_\mathrm{pk}$ and 
        DEC$_\mathrm{pk}$ are the J2000 coordinates of the NOEMA dust continuum image peaks; 
		$\nu_\mathrm{obs}$ is the central observed frequency. The rest-frame frequencies 
		of para-\hto\ \t202111, \t211202 and ortho-\hto\ \t321312 lines are: 987.927\,GHz, 
		752.033\,GHz and 1162.912\,GHz, respectively (the rest-frame frequencies are taken from the JPL catalogue: \url{http://spec.jpl.nasa.gov}); $t_\mathrm{on}$ is the on-source 
		integration time. The source G09v1.124, which is not resolved by SPIRE, is a cluster that
		consists of two main components: eastern component W (G09v1.124-W) and western component 
		T (G09v1.124-T) as described in \cite{2013ApJ...772..137I} (see also Fig.\,\ref{fig:map-3}).
}
\label{table:obs_log}
\end{table*}
\normalsize

\citetalias{2013ApJ...779...25B} provide lensing models, magnification 
factors ($\mu$) and inferred intrinsic properties of these galaxies and 
list their CO redshifts which come from \citet{2012ApJ...752..152H}; 
Harris et al. (in prep.); Lupu et al. (in prep.); Krips et al. (in prep.) 
and Riechers et al. (in prep.).

In our final selection of the sample to be studied in the \htot321312 
line, we then removed two sources, SDP\,81 and G12v2.30, that were 
previously observed in \hto\ (\citetalias{2013A&A...551A.115O}, and 
also \citealt{2015ApJ...808L...4A} for SDP\,81), because the $J=2$ 
\hto\ emission is too weak and/or the interferometry could resolve 
out some flux considering the lensing image. The observed \hz\ sample 
thus consists of two GAMA-field sources: G09v1.97 and G12v2.43, and 
four sources in the {\it H}-ATLAS NGP field: NCv1.143, NAv1.195, 
NAv1.177 and NBv1.78 (Tables\,\ref{table:obs_log} and 
\ref{table:previous_obs_properties}). Among the six remaining 
sources at redshift between 2.8 and 3.6, only one, NBv1.78, has 
been observed previously in a low-excitation line, para-\htot202111 
\citepalias{2013A&A...551A.115O}. Therefore, we have observed both para-\hto\ 
line \t202111 or \t211202 and ortho-\htot321312 in the other five sources, 
in order to compare their velocity-integrated flux densities.

\begin{table*}[htbp]
\setlength{\tabcolsep}{2.8pt}
\small
\caption{Previously observed properties of the sample.}
\centering
\begin{tabular}{cllllcccrccc}
\hline
\hline
Source          &  $z$                   & $F_{250}$                  &  $F_{350}$                 &  $F_{500}$                 & $F_{880}$                     & $r_\mathrm{half}$ & $\Sigma_{\mathit{SFR}}$                       & $f_{1.4GHz}\;\;\;$&  \td     &  $\mu$         & $\mu$\lir\         \\
                &                        & (mJy)                      &  (mJy)                     &  (mJy)                     & (mJy)                         & (kpc)             & (10$^{3}\,$M$_\odot$\,yr$^{-1}$\,kpc$^{-2}$)  & (mJy)\;\;\;  &  (K)    &                & (10$^{13}$\,\lsun) \\
\hline                                                                                                                                                                                                                        
G09v1.97        &  3.634                 &  $260\pm7$                 &  $321\pm8$                 &  $269\pm9$                 & $85.5\pm4.0$                  & $0.85$     & $0.91\pm0.15$                                 &     $\pm0.15$ & $44\pm1$ &$\;\;6.9\pm0.6$ &    $15.3\pm4.3$  \\
G12v2.43        &  3.127                 &  $290\pm7$                 &  $295\pm8$                 &  $216\pm9$                 & $48.6\pm2.3$                  &     --     &            --                                 &     $\pm0.15$ &       -- &           --   &\;\;($8.3\pm1.7$) \\
NCv1.143        &  3.565                 &  $214\pm7$                 &  $291\pm8$                 &  $261\pm9$                 & $97.2\pm6.5$                  & $0.40$     & $2.08\pm0.77$                                 & $0.61\pm0.16$ & $40\pm1$ &   $11.3\pm1.7$ &    $12.8\pm4.3$  \\
NAv1.195        &  2.951                 &  $179\pm7$                 &  $279\pm8$                 &  $265\pm9$                 & $65.2\pm2.3$                  & $1.57$     & $0.21\pm0.04$                                 &     $\pm0.14$ & $36\pm1$ &$\;\;4.1\pm0.3$ & $\;\;7.4\pm2.0$  \\
NAv1.177        &  2.778                 &  $264\pm9$                 & $310\pm10$                 & $261\pm10$                 & $50.1\pm2.1$                  &     --     &            --                                 &     $\pm0.15$ &       -- &           --   &\;\;($5.5\pm1.1$) \\
NBv1.78         &  3.111                 &  $273\pm7$                 &  $282\pm8$                 &  $214\pm9$                 & $59.2\pm4.3$                  & $0.55$     & $1.09\pm1.41$                                 & $0.67\pm0.20$ & $43\pm1$ &   $13.0\pm1.5$ &    $10.7\pm3.9$  \\
\hline                                                                                                                                                                                     
G09v1.124-W$^a$ & \multirow{2}{*}{2.410} & \multirow{2}{*}{$242\pm7$} & \multirow{2}{*}{$293\pm8$} & \multirow{2}{*}{$231\pm9$} & \multirow{2}{*}{$50.0\pm3.5$} &    --      & --                                            &     $\pm0.15$ & $40\pm1$ &             1  & $\;\;3.3\pm0.3$  \\
G09v1.124-T$^a$ &                        &                            &                            &                            &                               &    --      & --                                            &     $\pm0.15$ & $36\pm1$ &    $1.5\pm0.2$ & $\;\;2.7\pm0.8$  \\
G09v1.40        &  2.089$^b$             &  $389\pm7$                 &  $381\pm8$                 &  $241\pm9$                 & $61.4\pm2.9$                  & $0.41$     & $0.77\pm0.30$                                 & $0.75\pm0.15$ & $36\pm1$ &   $15.3\pm3.5$ & $\;\;6.5\pm2.5$  \\
SDP11           &  1.786                 &  $417\pm6$                 &  $378\pm7$                 &  $232\pm8$                 & $30.6\pm2.4$                  & $0.89$     & $0.22\pm0.08$                                 & $0.66\pm0.14$ & $41\pm1$ &   $10.9\pm1.3$ & $\;\;6.2\pm1.9$  \\
NCv1.268        &  3.675                 &  $145\pm7$                 &  $201\pm8$                 &  $212\pm9$                 & $78.9\pm4.4$                  & $0.93$     & $0.31\pm0.14$                                 & $1.10\pm0.14$ & $39\pm1$ &   $11.0\pm1.0$ & $\;\;9.5\pm2.7$  \\
NAv1.56         &  2.301                 &  $481\pm9$                 & $484\pm13$                 & $344\pm11$                 & $73.1\pm2.4$                  & $1.50$     & $0.14\pm0.08$                                 & $1.12\pm0.27$ & $38\pm1$ &   $11.7\pm0.9$ &    $11.3\pm3.1$  \\
\hline
\end{tabular}
\tablefoot{
        $z$ is the redshift inferred from previous CO detection quoted by 
        \citetalias{2013ApJ...779...25B} (see the references therein); $F_{250}$, 
        $F_{350}$ and $F_{500}$ are the SPIRE flux densities at 250, 350 and 500\mum,
        respectively \citep{2011MNRAS.415..911P}; $F_{880}$ is the SMA flux density 
		at 880\mum; $r_\mathrm{half}$ and $\Sigma_{\rm{SFR}}$ are the intrinsic 
		half-light radius at 880\,$\mu$m and the lensing-corrected surface 
		$\mathit{SFR}$ (star formation rate) density (Section\,\ref{hto}); 
		$f_{1.4GHz}$ is the 1.4\,GHz band flux densities from the VLA FIRST survey; 
		\td\ is the cold-dust temperature taken from \citetalias{2013ApJ...779...25B} 
		(note that the errors quoted for \td\ are significantly underestimated 
		since the uncertainties from differential lensing and single-temperature
		dust SED assumption are not fully considered); $\mu$ is the the lensing 
		magnification factor from \citetalias{2013ApJ...779...25B}, except for 
		G09v1.124 which is adopted from \cite{2013ApJ...772..137I}; 
		$\mu$\lir\ is the apparent total \ir\ luminosity mostly inferred from 
		\citetalias{2013ApJ...779...25B}. The $\mu$\lir\ in brackets are not listed in 
		\citetalias{2013ApJ...779...25B}, thus we infer them from single modified black 
		body dust SED fitting using the submm photometry data listed in this table. \\
		a: The cluster source G09v1.124 includes two main components: G09v1.124-W to 
		the east and G09v1.124-T to the west (Fig.\,\ref{fig:map-3}) and the values of 
		these two rows are quoted from \cite{2013ApJ...772..137I}; 
		b: Our \hto\ observation gives $z=2.093$ for G09v1.40. This value is slightly
		different from the value of 2.089 quoted by \citetalias{2013ApJ...779...25B} 
		from Lupu et al. (in prep.) obtained by CSO/Z-Spec, but consistent with \co32 
		observation by Riechers et al. (in prep.). 
}
\label{table:previous_obs_properties}
\end{table*}
\normalsize

In addition, we also observed five sources mostly at lower redshifts 
in para-\hto\ lines \t202111 or \t211202 (Tables\,\ref{table:obs_log} 
and \ref{table:previous_obs_properties}) to complete the sample of 
our \hto\ low-excitation study. They are three strongly lensed sources, 
G09v1.40, NAv1.56 and SDP11, a hyper-luminous cluster source G09v1.124 
\citep{2013ApJ...772..137I}, and a $z \sim 3.7$ source, NCv1.268 for 
which we did not propose a $J=3$ \hto\ observation, considering its 
large linewidth which could bring difficulties in line detection.

As our primary goal is to obtain a detection of the submm \hto\ lines, 
we carried out the observations in the compact, D configuration of NOEMA. 
The baselines extended from 24 to 176\,m, resulting in a synthesised 
beam with modest/low resolution of $\sim$\,$1.0$\,$''\times0.9$\,$''$ to 
$\sim$\,$5.6$\,$''\times3.3$\,$''$ as shown in Table \ref{table:obs_log}. The 
\hto\ observations were conducted from January 2012 to December 2013 in 
good atmospheric conditions (seeing of $0.3$\,$''$--$1.5$\,$''$) stability and 
reasonable transparency (PWV $\leq\,1\,\mathrm{mm}$). The total on source 
time was $\sim$\,1.5--8 hours per source. 2\,mm, 1.3\,mm and 0.8\,mm bands 
covering 129--174, 201--267 and 277--371\,GHz, respectively, were used. 
All the central observation frequencies were chosen based on previous 
redshifts given by \citetalias{2013ApJ...779...25B} according to the 
previous CO detections (Table\,\ref{table:previous_obs_properties}). 
In all cases but one, the frequencies of our detections of \hto\ 
lines are consistent with these CO redshifts. The only exception is 
G09v1.40 where our \hto\ redshift disagrees with the redshift of
$z=2.0894\pm0.0009$ given by Lupu et al. (in prep.), which is quoted 
by \citetalias{2013ApJ...779...25B}. We find $z=2.0925\pm0.0001$ in 
agreement with previous \co32 observations (Riechers et al., in prep.). 
We used the WideX correlator which provided a contiguous frequency 
coverage of 3.6\,GHz in dual polarisation with a fixed channel 
spacing of 1.95\,MHz.

The phase and bandpass were calibrated by measuring standard calibrators 
that are regularly monitored at the IRAM/NOEMA, including 3C279, 3C273, 
MWC349 and 0923+392. The accuracy of the flux calibration is estimated 
to range from $\sim$10\% in the 2\,mm band to $\sim$20\% in the 0.8mm 
band. Calibration, imaging, cleaning and spectra extraction were 
performed within the 
\texttt{GILDAS}\footnote{See \url{http://www.iram.fr/IRAMFR/GILDAS} 
for more information about the GILDAS softwares.} packages 
\texttt{CLIC} and \texttt{MAPPING}.

\begin{table}[htbp]
\small
\setlength{\tabcolsep}{1.45em}
\caption{Observed CO line properties using the IRAM 30m/EMIR.}
\centering
\begin{tabular}{lcccc}
\hline
\hline
Source     & CO line  & \ico               & \wco           \\
           &          & (Jy\,km\,s$^{-1}$) & (km\,s$^{-1}$) \\
\hline                                                       
G09v1.97   & \tco54   & $\;\;9.5\pm1.2$    & $224\pm 32$    \\
           & \tco65   & $10.4\pm2.3$       & $292\pm 86$    \\
NCv1.143   & \tco54   & $13.1\pm1.0$       & $273\pm 27$    \\
           & \tco65   & $11.0\pm1.0$       & $284\pm 27$    \\
NAv1.195   & \tco54   & $11.0\pm0.6$       & $281\pm 16$    \\
NAv1.177   & \tco32   & $\;\;6.8\pm0.4$    & $231\pm 15$    \\
           & \tco54   & $11.0\pm0.6$       & $230\pm 16$    \\
NBv1.78    & \tco54   & $10.3\pm0.8$       & $614\pm 53$    \\
           & \tco65   & $\;\;9.7\pm1.0$    & $734\pm 85$    \\
G09v1.40   & \tco43   & $\;\;7.5\pm2.1$    & $198\pm 51$    \\
NAv1.56    & \tco54   & $17.7\pm6.6$       & $\;\;432\pm 182$ \\
\hline
\end{tabular}
\tablefoot{
	       \ico\ is the velocity-integrated flux density of CO; \wco\ 
		   is the linewidth (FWHM) derived from fitting a 
		   single Gaussian to the line profile.
           }
\label{table:co_properties}
\end{table}
\normalsize

To compare the \hto\ emission with the typical molecular gas tracer, CO, we 
also observed the sources for CO lines using the EMIR receiver at the IRAM 
30m telescope. The CO data will be part of a systematic study of molecular 
gas excitation in {\it H}-ATLAS lensed Hy/ULIRGs, and a full description of 
the data and the scientific results will be given in a following paper 
(Yang et al., in prep.). The global CO emission properties of the sources 
are listed in Table\,\ref{table:co_properties} where we list the CO fluxes 
and linewidths. A brief comparison of the emission between \hto\ and CO 
lines will be given in Section \ref{CO lines}.

\section{Results}\label{Results}

A detailed discussion of the observation results for each source is given 
in Appendix\,\ref{Individual sources}, including the strength of the \hto\ 
emission, the image extension of \hto\ lines and the continuum 
(Fig.\,\ref{fig:map-all}), the \hto\ spectra and linewidths 
(Fig.\,\ref{fig:spectra-all}) and their comparison with CO 
(Table\,\ref{table:co_properties}). We give a synthesis of these results 
in this section.

\subsection{General properties of the \hto\ emissions}
\label{General properties}

 \begin{subfigures}
 \begin{figure*}[htbp]
	 \begin{center}
 \includegraphics[scale=0.85]{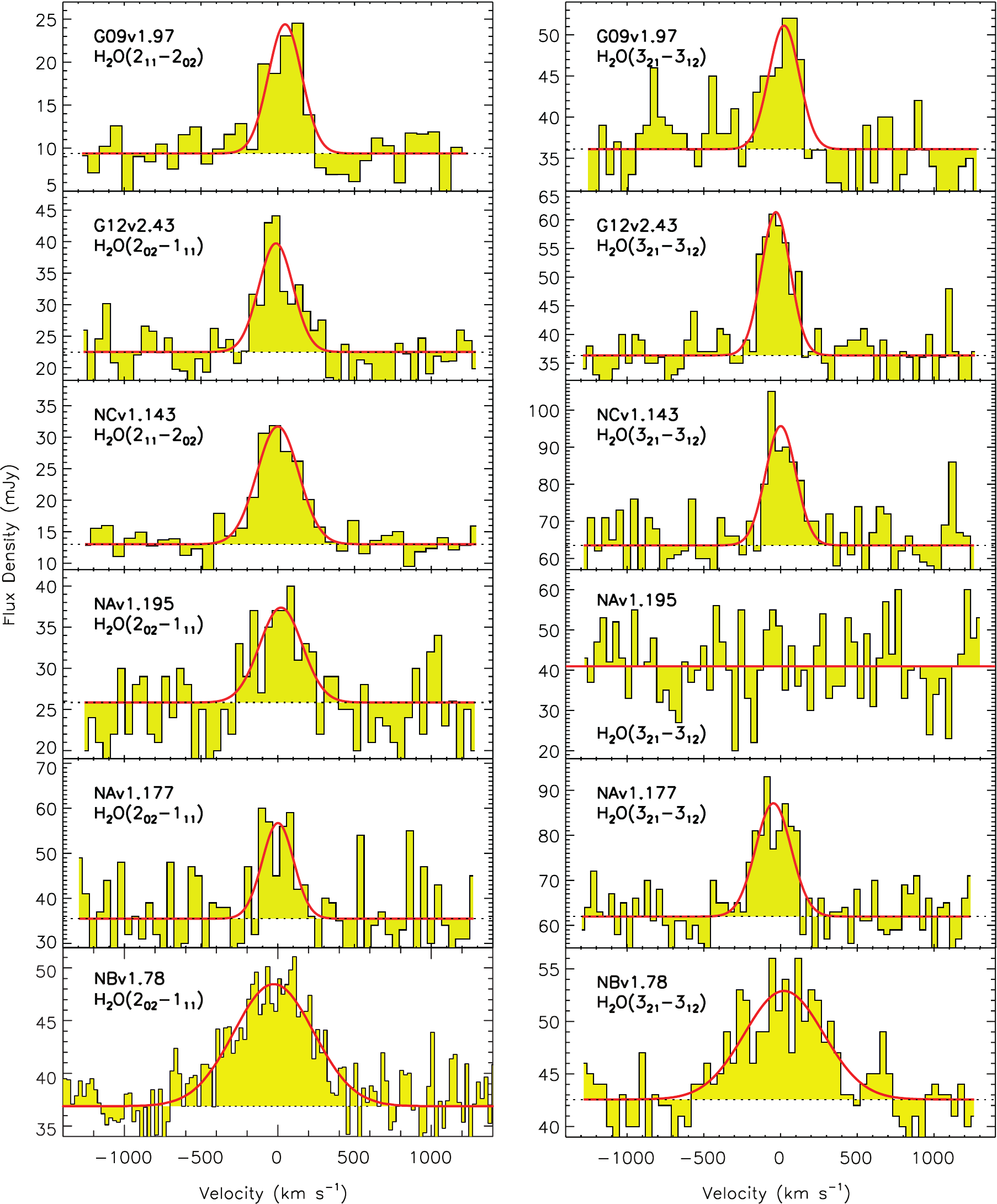}
 \caption{
		  Spatially integrated spectra of \hto\ in the six sources 
		  with both $J=2$ para-\hto\ and $J=3$ ortho-\hto\ lines observed. 
		  The red lines represent the Gaussian fitting to the emission 
		  lines. The \htot202111 spectrum of NBv1.78 is taken from 
		  \citetalias{2013A&A...551A.115O}. Except for \htot321312 in NAv1.195, 
		  all the $J=2$ and $J=3$ \hto\ lines are well detected, with a high 
		  S/N ratio and similar profiles in both lines for the same source.}
 \label{fig:spectra-1}
     \end{center}
 \end{figure*}

 \begin{figure*}[htbp]
	 \begin{center}
 \includegraphics[scale=0.97]{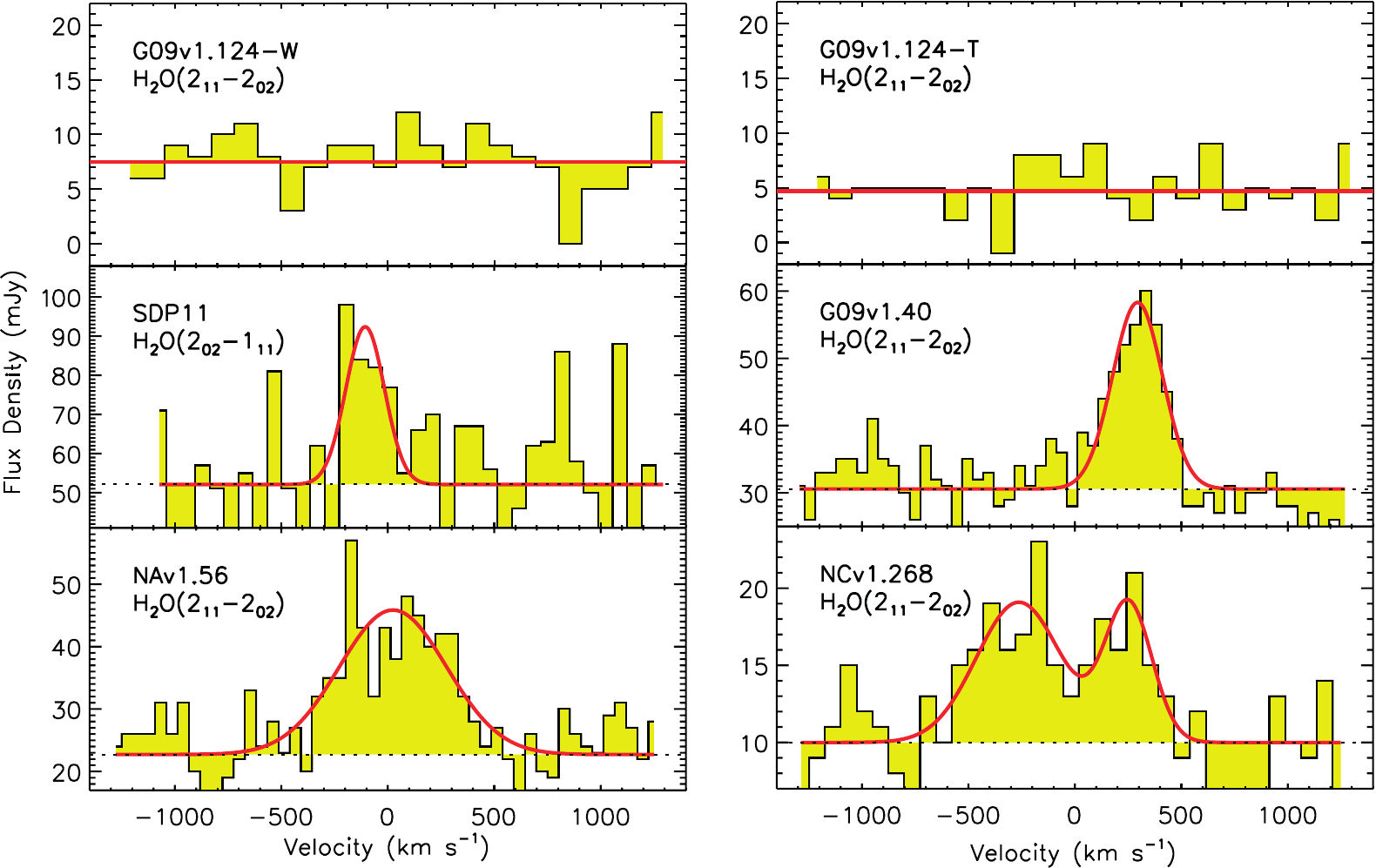}
 \caption{
 		  Spatially integrated spectra of \hto\ of the five sources 
 		  with only one $J=2$ para-\hto\ line observed. The red 
 		  lines represent the Gaussian fitting to the emission lines.
 		  Except for the \hto\ line in G09v1.124, 
 		  all the $J=2$ \hto\ lines are well detected.
		 }
 \label{fig:spectra-2}
     \end{center}
 \end{figure*}
 \label{fig:spectra-all}
\end{subfigures}


\begin{table*}[htbp]
\small
\setlength{\tabcolsep}{4.5pt}
\caption{The observed properties of \hto\ emission lines.}
\centering
\begin{tabular}{cccrrrrrrrr}
\hline
\hline
Source     & \hto\ line & $\nu_\mathrm{H_2O}$ & $S_{\nu}(\mathrm{ct})^{\mathrm{pk}}\:\:\:$ & $S_{\nu}(\mathrm{ct})\:\:\:\:\:$ & $S_\mathrm{H_2O}^{\mathrm{pk}}\:\:\:\:$ & $S_\mathrm{H_2O}\:\:\:\:$& \ihto$^{\mathrm{pk}}\:\:\:$           & \ihto\:\:\:\:\:\:    & \whto\:        & $\mu$\lhto\:\:\:\: \\
           &            & (GHz)               & ($\mathrm{mJy} \over \mathrm{beam}$)\:\:\:\:\:                   &  (mJy)\:\:\:\:\:               & ($\mathrm{mJy} \over \mathrm{beam}$)\:\:\:\:              & (mJy)\:\:\:\:  & ($\mathrm{Jy\,km\,s}^{-1} \over \mathrm{beam}$) & (Jy\,km\,s$^{-1}$) & (km\,s$^{-1}$) & (10$^8$\,\lsun)\:\:\:\\       
\hline                                                                                                                                                                                                                                    
G09v1.97   & \t211202   & 162.255             & $ 8.9\pm0.2$                       & $ 9.4\pm0.2$         & $14.9\pm2.2$                    &  $15.0\pm2.1$    & $ 3.8\pm0.4$                  & $ 4.1\pm0.4$     & $257\pm27$     & $ 7.4\pm0.7$  \\
           & \t321312   & 250.947             & $21.7\pm0.3$                       & $36.1\pm0.3$         & $ 7.8\pm1.9$                    &  $15.0\pm2.6$    & $ 2.4\pm0.4$                  & $ 3.7\pm0.4$     & $234\pm34$     & $10.4\pm1.0$  \\
G12v2.43   & \t202111   & 239.388             & $16.0\pm0.3$                       & $22.5\pm0.4$         & $10.8\pm2.1$                    &  $17.3\pm3.1$    & $ 3.2\pm0.5$                  & $ 4.8\pm0.6$     & $262\pm35$     & $ 8.8\pm1.0$  \\
           & \t321312   & 281.784             & $31.5\pm0.3$                       & $36.4\pm0.3$         & $25.6\pm3.3$                    &  $25.0\pm3.0$    & $ 4.9\pm0.4$                  & $ 5.9\pm0.5$     & $221\pm20$     & $12.7\pm1.0$  \\
NCv1.143   & \t211202   & 164.741             & $11.2\pm0.1$                       & $13.3\pm0.2$         & $17.4\pm1.3$                    &  $18.7\pm1.3$    & $ 5.6\pm0.3$                  & $ 5.8\pm0.3$     & $293\pm15$     & $10.1\pm0.5$  \\
           & \t321312   & 254.739             & $34.8\pm0.5$                       & $63.5\pm0.5$         & $23.9\pm4.3$                    &  $32.1\pm4.1$    & $ 5.2\pm0.6$                  & $ 8.0\pm0.7$     & $233\pm22$     & $21.3\pm1.8$  \\ 
NAv1.195   & \t202111   & 250.034             & $14.0\pm0.4$                       & $25.8\pm0.4$         & $ 6.6\pm2.5$                    &  $11.6\pm2.5$    & $ 2.1\pm0.6$                  & $ 4.0\pm0.6$     & $328\pm51$     & $ 6.7\pm1.0$  \\
           & \t321312   & (293.334)           & $17.2\pm0.5$                       & $41.2\pm0.5$         & $      <4.2$                    &  $      <7.3$    & $       <1.5$                 & $      <2.6$     & $ 330^{a}$     & $      <5.0$  \\
NAv1.177   & \t202111   & 261.489             & $26.5\pm0.6$                       & $35.5\pm0.6$         & $16.8\pm4.9$                    &  $21.2\pm4.9$    & $ 4.4\pm0.9$                  & $ 5.4\pm0.9$     & $241\pm41$     & $ 8.2\pm1.2$  \\
           & \t321312   & 307.856             & $38.2\pm0.4$                       & $62.0\pm0.4$         & $14.8\pm2.6$                    &  $25.2\pm3.1$    & $ 4.6\pm0.5$                  & $ 7.3\pm0.6$     & $272\pm24$     & $12.9\pm1.1$  \\
NBv1.78    & \;\,\t202111$^{b}$  & 240.290    & $15.4\pm0.3$                   & $36.9\pm0.4$     & $ 5.0\pm1.0$                    &  $12.3\pm3.2$    & $2.7\pm0.3$               & $6.7\pm1.3$  & $510\pm90$     & $12.2\pm2.4$\\
           & \t321312   & 282.863             & $29.2\pm0.2$                       & $42.6\pm0.2$         & $ 8.8\pm1.0$                    &  $10.6\pm1.0$    & $ 4.8\pm0.4$                  & $ 6.7\pm0.5$     & $607\pm43$     & $14.3\pm1.0$  \\
\hline                                                                                                                                                                                                
G09v1.124-W & \multirow{2}{*}{\t211202} & \multirow{2}{*}{(220.537)} & $6.42\pm0.15$& $7.6\pm0.2$         & $      <1.4$                    &  $      <1.6$    & $   <1.2^{c}$                 & $   <1.4^{c}$     & $850^{c}$      & $<1.3^{c}$    \\
G09v1.124-T &                           &                            & $4.08\pm0.15$& $4.9\pm0.2$         & $      <1.7$                    &  $      <2.0$    & $   <1.0^{c}$                 & $   <1.2^{c}$     & $550^{c}$      & $<1.0^{c}$    \\ 
G09v1.40   & \t211202   & 243.182             & $16.9\pm0.2$                       & $30.6\pm0.3$         & $17.5\pm2.0$                    &  $27.7\pm1.9$    & $ 4.9\pm0.4$                  & $ 8.2\pm0.4$      & $277\pm14$     & $ 5.7\pm0.3$  \\
SDP11      & \t202111   & 354.930             & $29.2\pm1.3$                       & $52.1\pm1.3$         & $14.8\pm8.4$                    &  $40.3\pm11.7$   & $ 5.2\pm2.0$                  & $ 9.2\pm2.0$      & $214\pm41$     & $ 6.3\pm1.1$  \\
NCv1.268   & \t211202   & 161.013             & $ 6.6\pm0.1$                       & $10.0\pm0.1$         & $ 5.2\pm1.1$                    &  $ 9.0\pm1.2$    & $ 3.7\pm0.4$                  & $ 7.0\pm0.7$      & $731\pm75$     & $12.8\pm1.2$  \\
NAv1.56    & \t211202   & 227.822             & $14.0\pm0.6$                       & $22.7\pm0.6$         & $15.8\pm3.3$                    &  $23.2\pm3.0$    & $ 7.8\pm1.1$                  & $14.6\pm1.3$      & $593\pm56$     & $12.0\pm1.1$  \\

\hline
\end{tabular}
\tablefoot{
		 $\nu_\mathrm{H_2O}$ is the observed central frequency of \hto\ 
		 lines, and the values in brackets are the \hto\ line frequencies 
		 inferred from the CO redshifts for the undetected sources; 
		 $S_{\nu}(ct)^{\mathrm{pk}}$ and $S_{\nu}(ct)$ are the peak and 
		 spatially integrated continuum flux density, respectively; 
		 $S_\mathrm{H_2O}^{\mathrm{pk}}$ is the peak \hto\ line flux and 
		 $S_\mathrm{H_2O}$ is the total line flux; \ihto$^{\mathrm{pk}}$ 
		 and \ihto\ are the peak and spatially integrated velocity-integrated 
		 flux density of the \hto\ lines; \whto\ is the \hto\ linewidth; 
		 $\mu$\lhto\ is the apparent luminosity of the observed \hto\ line. \\
		 a: The linewidth of the undetected \htot321312 in NAv1.195 has 
		    been set to 330\,km\,s$^{-1}$ by assuming that the widths of 
		    the \htot321312 and \htot202111 lines are roughly the same; 
		 b: The data of para-\htot202111 in NBv1.78 is taken from 
		    \citetalias{2013A&A...551A.115O};
		 c: the 2\,$\sigma$ upper limits of \ihto\ are derived by 
		    assuming that the \hto\ linewidths are similar to those 
		    of the CO lines \citep{2013ApJ...772..137I}.\\
 }
\label{table:h2o_properties}
\end{table*}
\normalsize

To measure the linewidth, velocity-integrated flux density and the 
continuum level of the spectra from the source peak and from the 
entire source, we extract each spectrum from the CLEANed image at the 
position of the source peak in a single synthesis beam and the 
spectrum integrated over the entire source. Then we fit them with 
Gaussian profiles using \texttt{MPFIT} \citep{2009ASPC..411..251M}.

We detect the high-excitation ortho-\htot321312 in five out of six observed 
sources, with high signal to noise ratios ($S/N\;>9$) and velocity-integrated 
flux densities comparable to those of the low-excitation $J=2$ para-\hto\ lines 
(Table \ref{table:h2o_properties} and Figs.\,\ref{fig:spectra-all} \& \ref{fig:map-all}). 
We also detect nine out of eleven $J=2$ para-\hto\ lines, either \t202111 or 
\t211202, with $S/N\,\ge\,6$ in terms of their velocity-integrated flux density, 
plus one 
tentative detection of \htot202111 in SDP11. We present the values of 
velocity-integrated \hto\ flux density detected at the source peak in a single synthesised beam, 
\ihto$^{\mathrm{pk}}$, and the velocity-integrated \hto\ flux density over the entire 
source, $I_{\mathrm{H_2O}}$ (Table\,\ref{table:h2o_properties}). The detected 
\hto\ lines are strong, with \ihto\;$= 3.7\text{--}14.6$\,Jy\,km\,s$^{-1}$. 
Even considering gravitational lensing correction, this is consistent with our 
previous finding that \hz\ Hy/ULIRGs are very strong \hto\ emitters, with \hto\ 
flux density approaching that of CO (Tables\,\ref{table:co_properties} \& 
\ref{table:h2o_properties} and Section\,\ref{CO lines}). The majority of the 
images (7/11 for $J=2$ lines and 3/4 for $J=3$) are marginally resolved with 
\ihto$^{\mathrm{pk}}$/$I_{\mathrm{H_2O}} \sim 0.4\text{--}0.7$. They show 
somewhat lensed structures. The others are unresolved with 
\ihto$^{\mathrm{pk}}$/$I_{\mathrm{H_2O}} > 0.8$. All continuum emission 
flux densities ($S_{\nu}(\mathrm{ct})^{\mathrm{pk}}$ for the emission peak 
and $S_{\nu}(\mathrm{ct})$ for the entire source) are very well detected 
($S/N \ge 30$), with a range of total flux density of 9--64\,mJy for 
$S_{\nu}(\mathrm{ct})$.
Fig.\,\ref{fig:map-all} shows the low-resolution images of \hto\ and 
the corresponding dust continuum emission at the observing frequencies. 
Because the positions of the sources were derived from {\it Herschel} 
observation, which has a large beamsize ($>17$\,$''$) comparing to the 
source size, the position of most of the sources are not perfectly centred 
at these {\it Herschel} positions as seen in the maps. The offsets are 
all within the position error of the {\it Herschel} measurement 
(Fig.\,\ref{fig:map-all}). G09v1.124 is a complex HyLIRG system 
including two main components eastern G09v1.124-W and western 
G09v1.124-T as described in \cite{2013ApJ...772..137I}. 
In Fig.\,\ref{fig:map-3}, we identified the two strong components 
separated about 10$''$, in agreement with \citet{2013ApJ...772..137I}. 
The $J=2$ \hto\ and dust continuum emissions in NBv1.78, NCv1.195, 
G09v1.40, SDP\,11 and NAv1.56, as well as the $J=3$ ortho-\hto\ and 
the corresponding dust continuum emissions in G09v1.97, NCv1.143 and 
NAv1.177, are marginally resolved as shown in Fig.\,\ref{fig:map-all}. 
Their images are consistent with the corresponding SMA images 
\citepalias{2013ApJ...779...25B} in terms of their spatial distribution. 
The rest of the sources are not resolved by the low-resolution 
synthesised beams. The morphological structure of the \hto\ emission 
is similar to the continuum for most sources as shown in 
Fig.\,\ref{fig:map-all}.
The ratio $S_{\nu}(\mathrm{ct})^{\mathrm{pk}}$/$S_{\nu}(\mathrm{ct})$ 
and $S_{\nu}(\mathrm{H_2O})^{\mathrm{pk}}$/$S_{\nu}(\mathrm{H_2O})$ 
are in good agreement within the error. However, for NCv1.143 in which 
$S_{\nu}(\mathrm{ct})^{\mathrm{pk}}$/$S_{\nu}(\mathrm{ct})=0.55\pm0.01$
and $S_{\nu}(\mathrm{H_2O})^{\mathrm{pk}}$/$S_{\nu}(\mathrm{H_2O})=0.74\pm0.16$, 
the $J=3$ ortho-\hto\ emission appears more compact than the dust continuum.
Generally it seems unlikely that we have a significant fraction of missing 
flux for our sources. Nevertheless, the low angular resolution ($\sim$\,$1''$ at best) 
limits the study of spatial distribution of the gas and dust in our 
sources. A detailed analysis of the images for each source is given 
in Appendix\,\ref{Individual sources}.

The majority of the sources have \hto\ (and CO) linewidths between 210 
and 330\,km\,s$^{-1}$, while the four others range between 500 and 
700\,km\,s$^{-1}$ (Table\,\ref{table:h2o_properties}). Except NCv1.268, 
which shows a double-peaked line profile, all \hto\ lines are well fit 
by a single Gaussian profile (Fig.\,\ref{fig:spectra-all}). The line 
profiles between the $J=2$ and $J=3$ \hto\ lines do not seem to be 
significantly different, as shown from the linewidth ratios ranging 
from $1.26\pm0.14$ to $0.84\pm0.16$. The magnification from strong 
lensing is very sensitive to the spatial configuration, in other words,
differential lensing, which could lead to different line profiles if the different 
velocity components of the line are emitted at different spatial positions. 
Since there is no visible differential effect between their profiles, 
it is possible that the $J=2$ and $J=3$ \hto\ lines are from similar 
spatial regions.

In addition to \hto, within the 3.6\,GHz WideX band, we have also 
tentatively detected \htop\ emission in 3 sources: NCv1.143, G09v1.97 
and G15v2.779 (see Section \ref{htop}).

\subsection{Lensing properties}
\label{Lensing properties}
All our sources are strongly gravitationally lensed (except G09v1.124, 
see Appendix\,\ref{g09v1.124}), which increases the line flux densities and
allows us to study the \hto\ emission in an affordable amount of observation 
time. However, the complexity of the lensed images complicates the analysis. 
As mentioned above, most of our lensed images are either unresolved or 
marginally resolved. Thus, we will not discuss here the spatial distribution 
of the \hto\ and dust emissions through gravitational lensing modelling. 
However, we should keep in mind that the correction of the magnification 
is a crucial part of our study. In addition, differential lensing could have 
a significant influence when comparing \hto\ emission with dust and even 
comparing different transitions of same molecular species 
\citep{2012MNRAS.424.2429S}, especially for the emission from close 
to the caustics.

In order to infer the intrinsic properties of our sample, especially 
\lhto\ as in our first paper \citetalias{2013A&A...551A.115O}, we adopted 
the lensing magnification factors $\mu$ 
(Table\,\ref{table:previous_obs_properties}) computed from the modelling
of the 880\,$\mu$m SMA images \citepalias{2013ApJ...779...25B}. As shown 
in the Appendix, the ratio of 
$S_{\nu}(\mathrm{ct})^{\mathrm{pk}}$/$S_{\nu}(\mathrm{ct})$ 
and $S_{\nu}(\mathrm{H_2O})^{\mathrm{pk}}$/$S_{\nu}(\mathrm{H_2O})$ 
are in good agreement within the uncertainties. Therefore, it is unlikely 
that the magnification of the 880\,$\mu$m continuum image and \hto\ 
can be significantly different. However, \citetalias{2013ApJ...779...25B} 
were unable to provide a lensing model for two of our sources, G12v2.43 
and NAv1.177, because their lens deflector is unidentified. 
This does not affect the modelling of \hto\ excitation and the 
comparison of \hto\ and \ir\ luminosities since the differential
lensing effect seems to be insignificant as discussed in 
Sections\,\ref{Discussion} and Appendix\,\ref{Individual sources}.

\begin{table*}[!htbp]
\small
\setlength{\tabcolsep}{1.5em}
\caption{IR luminosity, \hto\ line luminosity and global dust temperature of the entire sample.}
\centering
\begin{tabular}{cllccc}
\hline
\hline
Source     & \hto\ Transition   & \lir\               & \lhtot211202       & \lhtot202111        & \lhtot321312         \\
           &                    & ($10^{12}$\,\lsun)  & ($10^{7}$\,\lsun)  & ($10^{7}$\,\lsun)   & ($10^{7}$\,\lsun)    \\       
\hline                                                                                                                 
G09v1.97   & \t211202, \t321312 & $22.1\pm5.9$      & $10.7\pm1.4$     & --                  & $ 15.0\pm1.9$      \\
G12v2.43   & \t202111, \t321312 & $83.2\pm16.6/\mu$ & --                 & $88.4\pm10.7/\mu$ & $143.2\pm11.5/\mu$ \\
NCv1.143   & \t211202, \t321312 & $11.4\pm3.1$      & $ 9.0\pm1.4$     & --                  & $ 18.9\pm3.3$      \\
NAv1.195   & \t202111, \t321312 & $18.0\pm4.6$      & --                 & $16.4\pm3.0$      & $       <12.3$      \\
NAv1.177   & \t202111, \t321312 & $55.0\pm11.0/\mu$ & --                 & $82.0\pm12.8/\mu$ & $129.1\pm10.8/\mu$ \\
NBv1.78    & \t202111, \t321312 & $\;\;8.2\pm2.2$   & --                 & $ 9.4\pm2.1$      & $11.0\pm1.5$       \\
\hline
G09v1.124-W & \t211202          & $33.1\pm3.2$      & $<12.9$           & --                  & --                   \\
G09v1.124-T & \t211202          & $14.5\pm1.8$      & $<6.9$            & --                  & --                   \\
G09v1.40   & \t211202           & $\;\;4.2\pm1.3$   & $ 3.7\pm0.9$      & --                  & --                   \\
SDP11      & \t202111           & $\;\;5.7\pm1.6$   & --                 & $ 5.8\pm1.4$      & --                   \\
NCv1.268   & \t211202           & $\;\;8.6\pm2.3$   & $11.5\pm1.5$     & --                  & --                   \\
NAv1.56    & \t211202           & $\;\;9.7\pm2.6$   & $10.3\pm1.2$     & --                  & --                   \\
\hline                                                                                                                 
SDP81      & \t202111           & \;\;6.1             & --                 & 3.3                 & --                   \\
NAv1.144   & \t211202           & 11                  &  9.7               & --                  & --                   \\
SDP9       & \t211202           & \;\;5.2             &  7.0               & --                  & --                   \\
G12v2.30   & \t202111           & 16                  & --                 & 13                  & --                   \\
SDP17b     & \t202111           & 16                  & --                 & 20                  & --                   \\
G15v2.779  & \t211202           & 21                  &  26.6              & --                  & --                   \\
\hline
\end{tabular}
\tablefoot{
        $L_\mathrm{IR}$ is the intrinsic total \ir\ luminosity (8-1000\mum) taken from 
		\citetalias{2013ApJ...779...25B}. The intrinsic \hto\ luminosities are 
		inferred from $\mu$\lhto\ using $\mu$ in \citetalias{2013ApJ...779...25B}. The first 
		group of the sources are the ones with both $J=2$ and $J=3$ \hto\ lines observed, the 
		next group are the sources with only $J=2$ \hto\ observed, and the last group are the 
		previous published sources in \citetalias{2013A&A...551A.115O}.
		}
\label{table:Lir_Lh2o}
\end{table*}
\normalsize


\section{Discussion}
\label{Discussion}

\subsection{\lhto-\lir\ correlation and \lhto/\lir\ ratio}
\label{correlation}

Using the formula given by \cite{1992ApJ...387L..55S}, we derive the 
apparent \hto\ luminosities of the sources, $\mu$\lhto\ (Table\,
\ref{table:h2o_properties}), from \ihto. For the ortho-\htot321312
lines, $\mu$\lhto\ varies in the range of $6\text{--}22\times10^{8}$\,\lsun, 
while the $\mu$\lhto\ of the $J=2$ lines are a factor $\sim1.2\text{--}2$ 
weaker (Table\,\ref{table:h2o_properties}) as discussed in Section\,\ref{hto}.

\begin{figure*}[htbp]
	 \begin{center}
 \includegraphics[scale=0.62]{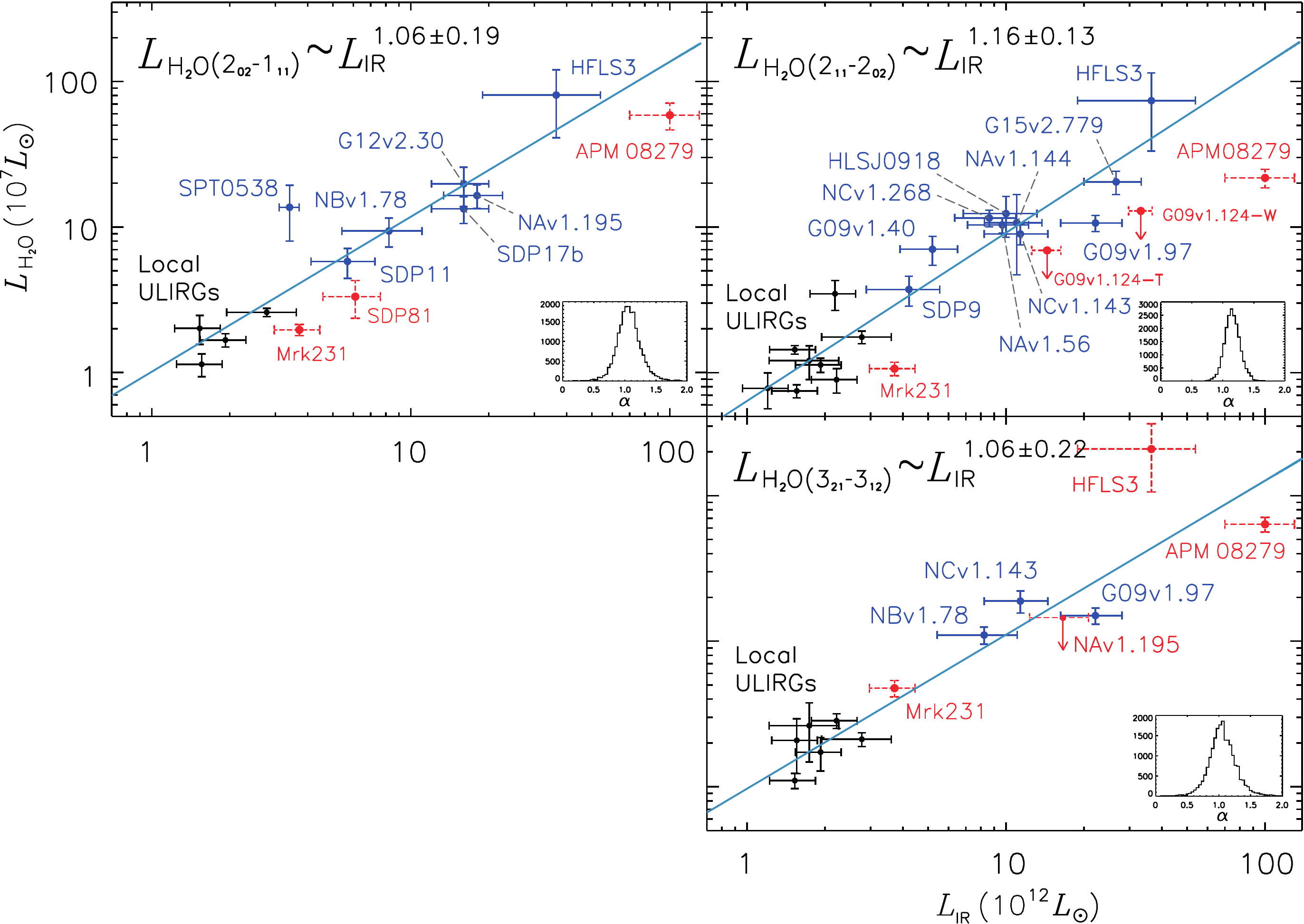}
 \caption{
		  Correlation between \lir\ and \lhto\ in local ULIRGs and \hz\ Hy/ULIRGs. 
		  The black points represent local ULIRGs from \citetalias{2013ApJ...771L..24Y}. 
		  The blue points with solid error bars are the {\it H}-ATLAS source in 
		  this work together with some previously published sources. Red points 
		  with dashed error bars are excluded from the fit as described in the text. 
		  Upper limits are shown in arrows. The light blue lines show the results 
		  of the fitting. The insets are the probability density distributions of 
		  the fitted slopes $\alpha$. We find tight correlations between the 
		  luminosity of the three \hto\ lines and \lir, namely \lhto\;$\propto$\;\lir$^{1.1-1.2}$.
         }
 \label{fig:h2o-ir}
     \end{center}
 \end{figure*}

Using the lensing magnification correction (taking the values of $\mu$ 
from \citetalias{2013ApJ...779...25B}), we have derived the intrinsic \hto\ 
luminosities (Table\,\ref{table:Lir_Lh2o}). The error of each luminosity 
consists of the uncertainty from both observation and the gravitational 
lensing modelling. After correcting for lensing, the \hto\ luminosities 
of our \hz\ galaxies appear to be one order of magnitude higher than those of 
local ULIRGs, as well as their \ir\ luminosities (Table\,\ref{table:Lir_Lh2o}), 
so that many of them should rather be considered as HyLIRGs than ULIRGs. 
Though the ratio of \lhto/\lir\ in our \hz\ sample is close to that of 
local ULIRGs \citepalias{2013ApJ...771L..24Y}, with somewhat a statistical 
increase in the extreme high \lir\ end (Fig.\,\ref{fig:h2o-ir}).

As displayed in Fig.\,\ref{fig:h2o-ir} for \hto\ of the three observed 
lines, because we have extended the number of detections to 21 \hto\ 
lines, distributed in 16 sources and 3 transitions, we may independently
study the correlation of \lhtot202111 and \lhtot211202 with \lir, while we
had approximately combined the two lines in \citetalias{2013A&A...551A.115O}.

As found in \citetalias{2013A&A...551A.115O}, the correlation is slightly 
steeper than linear (\lhto\;$\sim$\;\lir$^{1.2}$). To broaden the dynamical 
range of this comparison, we also included the local ULIRGs from 
\citetalias{2013ApJ...771L..24Y}, together with a few other \hto\ detections 
in \hz\ Hy/ULIRGs, for example, HLSJ\,0918 (HLSJ\,091828.6+514223) 
\citep{2012A&A...538L...4C, 2014ApJ...783...59R}, APM\,08279 
\citep{2011ApJ...741L..38V}, SPT\,0538 (SPT-S\,J053816−5030.8) 
\citep{2013ApJ...779...67B} and HFLS3 (\citealt{2013Natur.496..329R}, 
with the magnification factor from \citealt{2014ApJ...790...40C}) 
(Fig.\,\ref{fig:h2o-ir}). In the fitting, 
however, we excluded the sources with heavy AGN contamination (Mrk\,231 and 
APM\,08279) or missing flux resolved out by the interferometry (SDP\,81). We 
also excluded the \htot321312 line of HFLS3 considering its unusual high 
\lhtot321312/\lir\ ratio as discussed above, that could bias our fitting. 
We have performed a linear regression in log-log space using the 
Metropolis-Hastings Markov Chain Monte Carlo (MCMC) algorithm sampler 
through {\texttt{linmix\_err}} \citep{2007ApJ...665.1489K} to derived 
the $\alpha$ in
\begin{equation} 
	\label{eq:L_L}
L_\mathrm{H_{2}O} \propto L_\mathrm{IR}^\alpha.
\end{equation} 
The fitted parameters are $\alpha=1.06\pm0.19$, $1.16\pm0.13$ and 
$1.06\pm0.22$ for \hto\ line \t202111, \t211202 and \t321312, respectively. 
Comparing with the local ULIRGs, the \hz\ lensed ones have higher \lhto/\lir\ 
ratios (Table\,\ref{table:Lir_Lh2o_ratios}). These slopes confirm our first 
result derived from 7 \hto\ detections in \citepalias{2013A&A...551A.115O}. 
The slight super-linear correlations seem to indicate that \fir\ pumping 
play an important role in the excitation of the submm \hto\ emission. This 
is unlike the high-$J$ CO lines, which are determined by collisional 
excitation and follow the linear correlation between the CO line luminosity 
and \lir\ from the local to the \hz\ Universe \citep{2015ApJ...810L..14L}. 
As demonstrated in \citetalias{2014A&A...567A..91G}, using the \fir\ pumping 
model, the steeper than linear growth of \lhto\ with \lir\ can be the result 
of an increasing optical depth at 100\,$\mu$m ($\tau_{100}$) with increasing 
\lir. In local ULIRGs, the ratio of \lhto/\lir\ is relatively low while most 
of them are likely to be optically thin 
\citepalias[$\tau_{100}\sim0.1$,][]{2014A&A...567A..91G}. On the other hand, 
for the \hz\ lensed Hy/ULIRGs with high values of \lir, the continuum optical 
depth at \fir\ wavelengths is expected to be high (see Section\,\ref{hto}), 
indicating that the \hto\ emission comes from very dense regions of molecular 
gas that are heavily obscured.

\begin{table*}[!htbp]
\small
\setlength{\tabcolsep}{0.88em}
\caption{Ratio between \ir\ and \hto\ luminosity, and the velocity-integrated flux density ratio between different \hto\ transitions.}
\centering
\begin{tabular}{clcccccc}
\hline
\hline
Source     & \hto\ Transition   & \td       & \lhtotlir211202    & \lhtotlir202111    & \lhtotlir321312    & \ihtotihto211202 & \ihtotihto202111 \\
           &                    & (K)       & ($\times 10^{-6}$) & ($\times 10^{-6}$) & ($\times 10^{-6}$) &                  &                  \\
\hline                                                                                                                  
G09v1.97   & \t211202, \t321312 & $44\pm1$  & $ 4.8\pm1.4$       & --                 & $ 6.8\pm2.0$       & $0.9\pm0.1$      & ($0.8\pm0.2$)    \\
G12v2.43   & \t202111, \t321312 &($39\pm2$) & --                 & $10.6\pm2.5$       & $15.3\pm3.3$       & --               & $1.2\pm0.2$      \\
NCv1.143   & \t211202, \t321312 & $40\pm1$  & $ 7.9\pm2.5$       & --                 & $16.6\pm5.4$       & $1.4\pm0.1$      & ($1.1\pm0.4$)    \\
NAv1.195   & \t202111, \t321312 & $36\pm1$  & --                 & $ 9.1\pm2.9$       & $       <6.8$      & --               & $      <0.7$     \\
NAv1.177   & \t202111, \t321312 &($32\pm1$) & --                 & $14.9\pm3.8$       & $23.5\pm5.1$       & --               & $1.3\pm0.2$      \\
NBv1.78    & \t202111, \t321312 & $43\pm1$  & --                 & $11.4\pm4.7$       & $13.4\pm4.9$       & --               & $1.0\pm0.2$      \\ 
G09v1.124-W & \t211202          & $40\pm1$  & $<3.9$             & --                 & --                 & --               & --               \\
G09v1.124-T & \t211202          & $36\pm1$  & $<4.8$             & --                 & --                 & --               & --               \\
G09v1.40   & \t211202           & $36\pm1$  & $ 8.8\pm3.5$       & --                 & --                 & --               & --               \\
SDP11      & \t202111           & $41\pm1$  & --                 & $10.2\pm3.8$       & --                 & --               & --               \\
NCv1.268   & \t211202           & $39\pm1$  & $13.4\pm3.9$       & --                 & --                 & --               & --               \\
NAv1.56    & \t211202           & $38\pm1$  & $10.7\pm3.1$       & --                 & --                 & --               &                  \\
\hline                                     
SDP81      & \t202111           & $34\pm1$  & --                 & $ 5.4$             & --                 & --               & --               \\
NAv1.144   & \t211202           & $39\pm1$  & $ 9.7$             & --                 & --                 & --               & --               \\
SDP9       & \t211202           & $43\pm1$  & $13.5$             & --                 & --                 & --               & --               \\
G12v2.30   & \t202111           & $41\pm1$  & --                 & $ 8.1$             & --                 & --               & --               \\
SDP17b     & \t202111           & $38\pm1$  & --                 & $12.5$             & --                 & --               & --               \\
G15v2.779  & \t211202           & $41\pm1$  &  $7.7$             & --                 & --                 & --               & --               \\
\hline
HFLS3                      & \t202111, \t211202, \t321312 & $56^{+9}_{-12}$ & $20.3$  & $22.2$   & $57.3$             & $1.8\pm0.6$      & $2.2\pm0.5$      \\
APM\,08279                 & \t202111, \t211202, \t321312 &$220\pm30$ & $ 2.2$  & $ 6.0$   & $6.4$              & $1.9\pm0.3$      & $0.9\pm0.1$      \\
HLSJ\,0918                 & \t202111                     & $38\pm3$ & $11.4$  &  --      & --                 & --               & --               \\
SPT\,0538                  & \t202111                     & $39\pm2$ & --      & $40.3$   & --                 & --               & --               \\
\hline                                                           
local strong-AGN           & \t202111, \t211202, \t321312 & --   &  $3.8$  &  $6.4$   & $6.7$              & $1.1\pm0.4$      & $0.9\pm0.3$      \\
local \ion{H}{ii}+mild-AGN & \t202111, \t211202, \t321312 & --   &  $5.8$  &  $9.2$   & $10.8$             & $1.4\pm0.4$      & $1.1\pm0.3$      \\
\hline
\end{tabular}
\tablefoot{
		The luminosity ratios between each \hto\ line and their total \ir,
		and the velocity-integrated flux density ratio of different \hto\ transitions. \td\ is the cold-dust 
		temperature taken from \citetalias{2013ApJ...779...25B}, except for the 
		ones in brackets which are not listed \citetalias{2013ApJ...779...25B}, 
		that we infer them from single modified black-body dust SED fitting using 
		the submm/mm photometry data listed in Table\,\ref{table:previous_obs_properties}.
		All the errors quoted for \td\ are significantly underestimated especially 
		because they do not include possible effects of differential lensing and make 
		the assumption of a single-temperature. Line ratios in brackets are derived 
		based on the average velocity-integrated flux density ratios between \t211202 and \t202111 lines 
		in local infrared galaxies. The local strong-AGN sources are the optically 
		classified AGN-dominated galaxies and the local \ion{H}{ii}+mild-AGN sources 
		are star-forming-dominated galaxies with possible mild AGN contribution 
		\citepalias{2013ApJ...771L..24Y}. The first group of the sources are from 
		this work; and the sources in the second group are the previously published 
		sources in \citetalias{2013A&A...551A.115O}; the third group contains the 
		previously published \hz\ detections from other works: 
		HFLS3 \citep{2013Natur.496..329R}, APM\,08279 \citep{2011ApJ...741L..38V}, 
		HLSJ\,0918 \citep{2012A&A...538L...4C, 2014ApJ...783...59R} and SPT\,0538 \citep{2013ApJ...779...67B};
		the last group shows the local averaged values from \citetalias{2013ApJ...771L..24Y}.

 }
\label{table:Lir_Lh2o_ratios}
\end{table*}
\normalsize

Similar to what we found in the local ULIRGs \citepalias{2013ApJ...771L..24Y}, we find 
again an anti-correlation between \td\ and \lhtot321312/\lir. The Spearman$'$s rank 
correlation coefficient for the five \htot321312 detected {\it H}-ATLAS sources is 
$\rho = -0.9$ with a two-sided significance of its deviation from zero, $p=0.04$. 
However, after including the non-detection of \htot321312 in NAv1.195, the correlation 
is much weaker, that is to say, $\rho \lesssim -0.5$ and $p \sim 0.32$. No significant 
correlation has been found between \td\ and \lhtot202111/\lir\ ($\rho = -0.1$ 
and $p=0.87$) nor \lhtot211202/\lir\ ($\rho = -0.3$ and $p=0.45$). As explained 
in \citetalias{2014A&A...567A..91G}, in the optically thick and very warm galaxies, 
the ratio of \lhtot321312/\lir\ is expected to decrease with increasing \td. And this 
anti-correlation can not be explained by optically thin conditions. However, a larger 
sample is needed to increase the statistical significance of this anti-correlation.

Although, it is important to stress that the luminosity of \hto\ is a 
complex result of various physical parameters such as dust temperature, 
gas density, \hto\ abundance and \hto\ gas distribution relative to the \ir\ 
radiation field, etc, it is striking that the correlation between \lhto\ and 
\lir\ stays linear from local young stellar objects (YSOs), in which the \hto\ 
molecules are mainly excited by shocks and collisions, to local ULIRGs 
(\fir\ pumping dominated), extending $\sim12$ orders of magnitudes 
\citep{2016A&A...585A.103S}, implying that \hto\ indeed traces the SFR 
proportionally, similarly to the dense gas \citep{2004ApJ...606..271G} 
in the local infrared bright galaxies. However, for the \hz\ sources, 
the \lhto\ emissions are somewhat above the linear correlations which 
could be explained by their high $\tau_{100}$ (or large velocity dispersion). 
As shown in Table\,\ref{table:Lir_Lh2o_ratios}, HFLS3, with a $\tau_{100}>1$ 
has extremely large ratios of \lhto/\lir\ which are stronger than the 
average of our {\it H}-ATLAS sources by factors $\sim$\,2 for the $J=2$ 
lines and $\sim$\,4 for $J=3$ (see Fig.\,\ref{fig:h2o-ir}). The velocity 
dispersions of its \hto\ lines are $\sim$\,900\,km\,s$^{-1}$ (with 
uncertainties from 18\% to 36\%), which is larger than all our sources. 
For optically thick systems, larger velocity dispersion will increase 
the number of absorbed pumping photons, and boost the ratio of 
\lhto/\lir\ \citepalias{2014A&A...567A..91G}. 

For the AGN-dominated sources (i.e. APM\,08279, G09v1.124-W and 
Mrk\,231) as shown in Fig.\,\ref{fig:h2o-ir}, most of them (except 
for the \htot321312 line of Mrk\,231) are well below the fitted correlation 
(see Section\,\ref{AGN}). 
This is consistent with the average value of local strong-AGN-dominated 
sources. The $J \lesssim 3$ \hto\ lines are \fir\ pumped by the 75 
and 101\,$\mu m$ photons, thus the very warm dust in strong-AGN-dominated 
sources is likely to contribute more to the \lir\ than the $J \lesssim 3$ 
\hto\ excitation (see also \citetalias{2013ApJ...771L..24Y}).

\subsection{\hto\ excitation}\label{hto}
We have detected both $J=2$ and $J=3$ \hto\ lines in five sources 
out of six observed for $J=3$ ortho-\hto\ lines. By comparing the 
line ratios and their strength relative to \lir, we are able to  
constrain the physical conditions of the molecular content 
and also the properties of the \fir\ radiation field.

 \begin{figure}[htbp]
	 \begin{center}
 \includegraphics[scale=0.561]{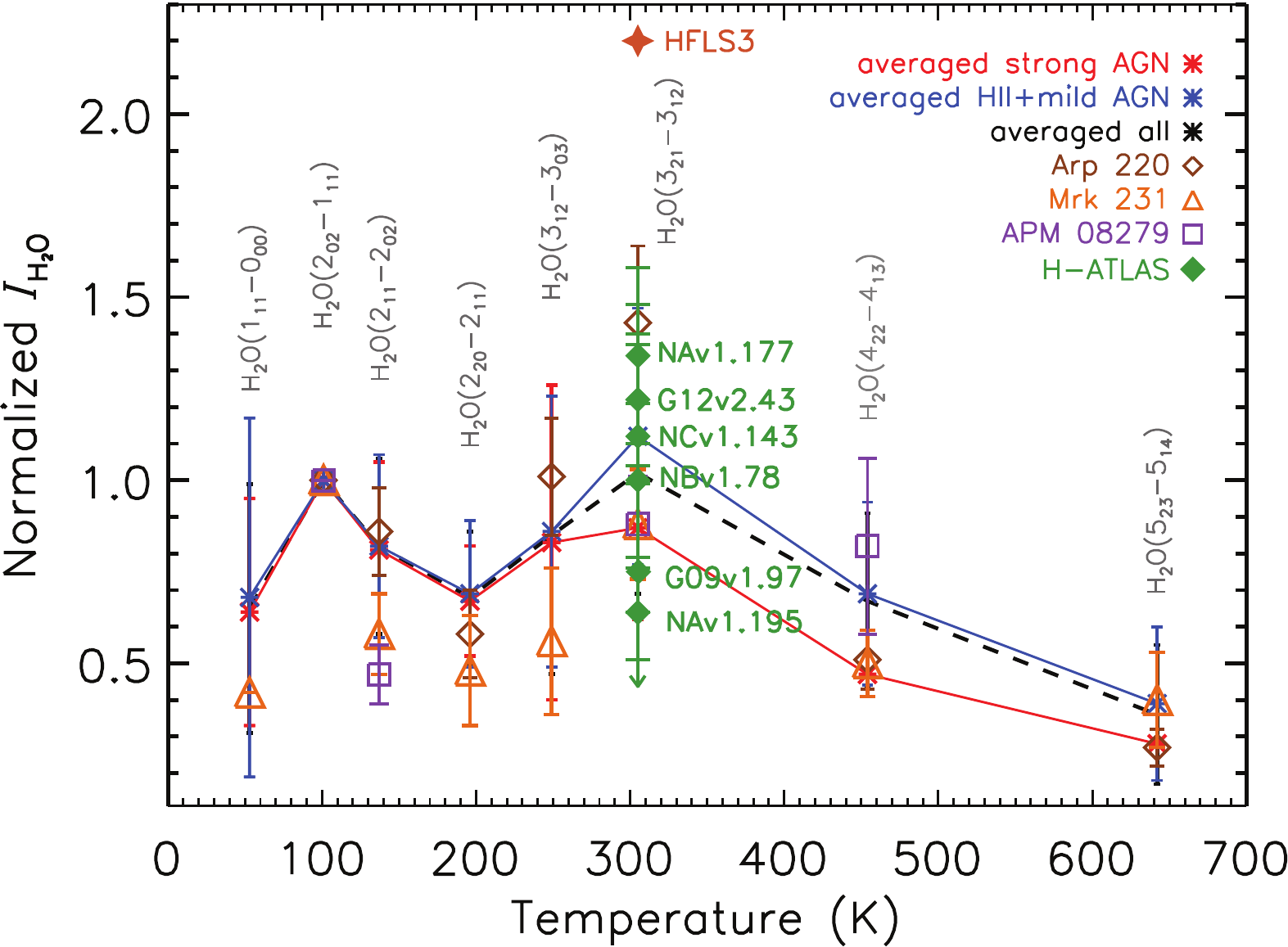}
 \caption{
		  Velocity-integrated flux density distribution of \hto\ normalised to 
		  \ihtot202111 adapted from \citetalias{2013ApJ...771L..24Y}. Local 
		  averaged values are shown in black dashed line and marks. Among them, 
		  AGN-dominated sources are shown in red and star-forming dominated galaxies 
		  are shown in blue. Some individual sources are also shown in this plot as
		  indicated by the legend. Green diamonds are the \hz\ lensed 
		  Hy/ULIRGs from this work. HFLS3 is a $z=6.3$ \hz\ galaxy 
		  from \cite{2013Natur.496..329R}.
		  }
 \label{fig:h2o-sled}
     \end{center}
 \end{figure}

\begin{figure*}[!htbp]
	 \begin{center}
 \includegraphics[scale=0.82]{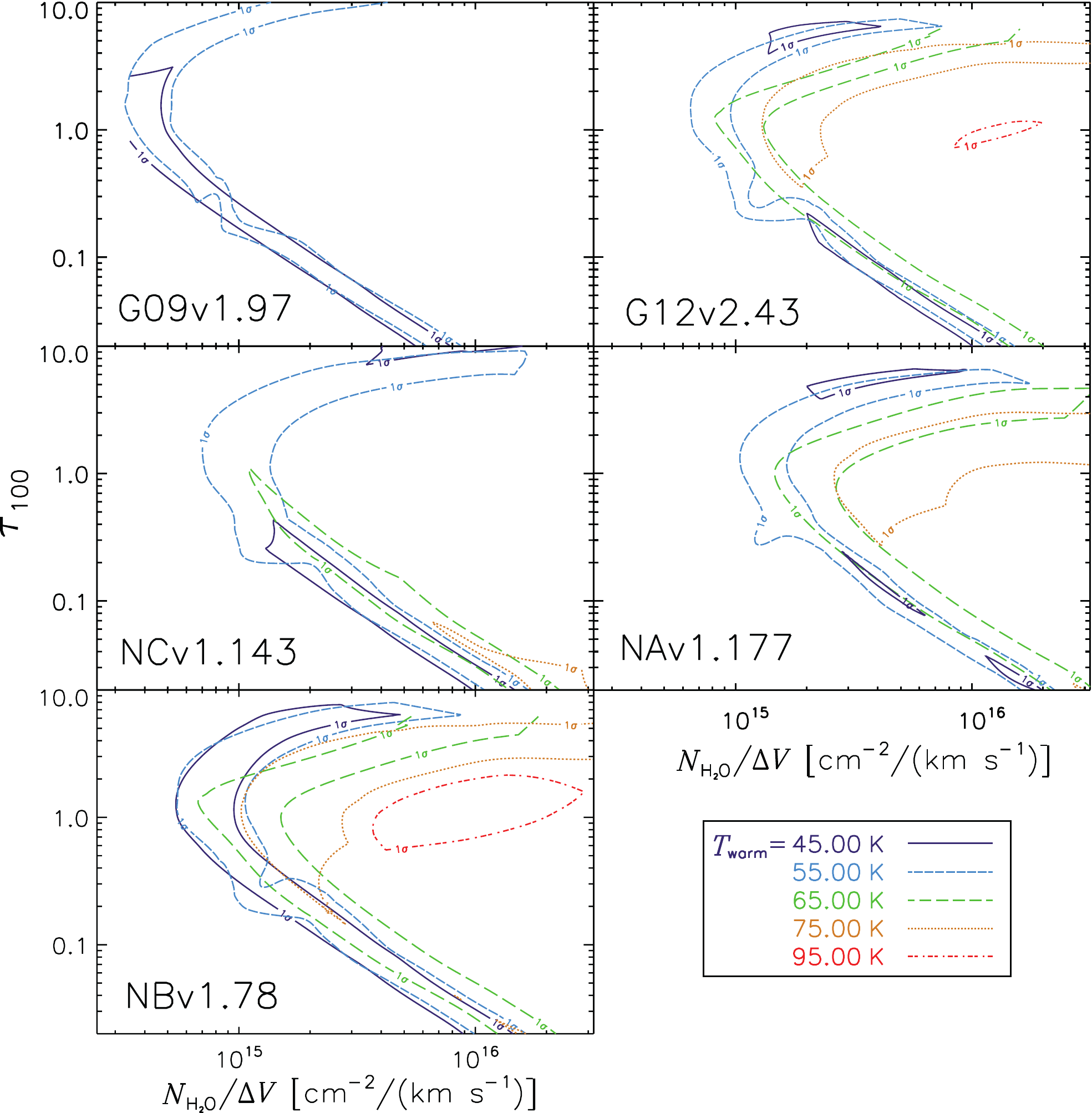}
 \caption{
		  Parameter space distribution of the H$_2$O \fir\ pumping excitation 
		  modelling with observed para-\hto\ \t202111 or \t211202 and 
		  ortho-\htot321312 in each panel. $\pm 1\,\sigma$ contours are 
		  shown for each plot. Different colours with different line styles 
		  represent different temperature components of the warm dust as 
		  shown in the legend. The explored warm dust temperature range is 
		  from 35\,K to 115\,K. The temperature contours that are unable to 
		  fit the data are not shown in this figure. From the figure, we 
		  are able to constrain the $\tau_{100}$, $T_\mathrm{warm}$ and
		  \nhto/$\Delta V$ for the five sources. However, there are strong
		  degeneracies. Thus, we need additional information, such as the 
		  velocity-integrated flux densities of $J\geq4$ \hto\ lines, to 
		  better constrain the physical parameters.  
		  }
 \label{fig:h2o-model}
     \end{center}
 \end{figure*}

To compare the \hto\ excitation with local galaxies, we plot the
velocity-integrated flux density of ortho-\htot321312 normalised by that 
of para-\htot202111 in our source on top of the local and \hz\ \hto\ 
SLEDs (spectral line energy distributions) in Fig.\,\ref{fig:h2o-sled}. 
All the six \hz\ sources are located within the range of the local galaxies, 
with a 1\,$\sigma$ dispersion of $\sim 0.2$. Yet for the $z=6.34$ 
extreme starburst HFLS3, the value of this ratio is at least 1.7 times
higher than the average value of local sources \citepalias{2013ApJ...771L..24Y}
and those of our lensed \hz\ Hy/ULIRGs at $\gtrapprox3\,\sigma$ confidence 
level (Fig.\,\ref{fig:h2o-sled}). This probably traces different 
excitation conditions, namely the properties of the dust emission, 
as it is suggested in \citetalias{2014A&A...567A..91G} that the flux
ratio of \htot321312 over \htot202111 is the most direct tracer 
of the hardness of the \fir\ radiation field which powers the submm
\hto\ excitation. However, the line ratios are still consistent 
with the strong saturation limit in the \fir\ pumping model with a 
$T_\mathrm{warm} \gtrsim 65$\,K. The large scatter of the \hto\ line 
ratio between \t321312 and \t202111 indicates different local \hto\ 
excitation conditions. As \fir\ pumping is dominating the \hto\ 
excitation, the ratio therefore reflects the differences in the \fir\ 
radiation field, for example, the temperature of the warmer dust that excites 
the \hto\ gas, and the submm continuum opacity. It is now clear that 
\fir\ pumping is the prevailing excitation mechanism for those submm 
\hto\ lines rather than collisional excitation 
\citepalias{2014A&A...567A..91G} in \ir\ bright galaxies in both the 
local and \hz\ Universe. The main path of \fir\ pumping related to the 
lines we observed here are 75 and 101\,$\mu$m as displayed in 
Fig.\,\ref{fig:h2o-e-level}. Therefore, the different line ratios are 
highly sensitive to the difference between the monochromatic flux at 
75 and 101\,$\mu$m. We may compare the global \td\ measured from \fir\ 
and submm bands \citepalias{2013ApJ...779...25B}. It includes both cold 
and warm dust contribution to the dust SED in the rest-frame, which is, 
however, dominated by cold dust observed in SPIRE bands. It is thus 
not surprising that we find no strong correlation between \td\ and 
\ihtot321312/\ihtot202111 ($r \sim -0.3$). The Rayleigh-Jeans tail of 
the dust SED is dominated by cooler dust which is associated with 
extended molecular gas and less connected to the submm \hto\ excitation. 
As suggested in \citetalias{2014A&A...567A..91G}, it is indeed the 
warmer dust ($T_\mathrm{warm}$, as shown by the colour legend in 
Fig.\,\ref{fig:h2o-model}) dominating at the Wien side of the dust 
SED that corresponds to the excitation of submm \hto\ lines.

To further explore the physical properties of the \hto\ gas content 
and the \fir\ dust radiation related to the submm \hto\ excitation, 
we need to model how we can infer key parameters, such as the \hto\ 
abundance and those determining the radiation properties, from the
observed \hto\ lines. For this purpose, we use the \fir\ pumping 
\hto\ excitation model described in
\citetalias{2014A&A...567A..91G} to fit the observed \lhto\ together 
with the corresponding \lir, and derive the range of continuum optical 
depth at 100\,$\mu$m ($\tau_{100}$), warm dust temperature 
($T_\mathrm{warm}$), and \hto\ column density per unit of velocity 
interval (\nhto/$\Delta V$) in the five sources with both $J=2$ and 
$J=3$ \hto\ emission detections. Due to the insufficient number of 
the inputs in the model, which are \lhto\ of the two \hto\ lines and \lir, 
we are only able to perform the modelling by using the pure \fir\ pumping 
regime. Nevertheless, our observed line ratio between $J=3$ and $J=2$ 
\hto\ lines suggests that \fir\ pumping is the dominant excitation 
mechanism and the contribution from collisional excitation is minor 
\citepalias{2014A&A...567A..91G}. The $\pm 1\,\sigma$ contours from 
$\chi^2$ fitting are shown in Fig.\,\ref{fig:h2o-model} for each warm 
dust temperature component ($T_\mathrm{warm} = 35\text{--}115$\,K) 
per source. It is clear that with two \hto\ lines (one $J=2$ para-\hto\ 
and ortho-\htot312312), we will not be able to well constrain $\tau_{100}$ 
and \nhto/$\Delta V$. As shown in the figure, for $T_\mathrm{warm} \lesssim 75$\,K, 
both very low and very high $\tau_{100}$ could fit the observation data together 
with high \nhto/$\Delta V$, while the dust with $T_\mathrm{warm} \gtrsim 95$\,K 
are likely favouring high $\tau_{100}$. In the low continuum optical depth part 
in Fig.\,\ref{fig:h2o-model}, as $\tau_{100}$ decreases, the model needs to 
increase the value of \nhto/$\Delta V$ to generate sufficient \lhto\ to be able 
to fit the observed \lhto/\lir. This has been observed in some local sources 
with low $\tau_{100}$, such as in NGC\,1068 and NGC\,6240. There are no absorption 
features in the \fir\ but submm \hto\ emission have been detected in these 
sources \citepalias{2014A&A...567A..91G}. The important feature of such 
sources is the lack of $J\geq4$ \hto\ emission lines. Thus, the observation 
of higher excitation of \hto\ will discriminate between the low and high 
$\tau_{100}$ regimes.

Among these five sources, favoured key parameters are somewhat different 
showing the range of properties we can expect for such sources. Compared 
with the other four Hy/ULIRGs, G09v1.97 is likely to have the lowest 
$T_\mathrm{warm}$ as only dust with $T_\mathrm{warm} \sim 45-55$\,K can 
fit well with the data. NCv1.143 and NAv1.177 have slightly different 
diagnostic which yields higher dust temperature as $T_\mathrm{warm} \sim 45\text{--}75$\,K, 
while NBv1.78 and G12v2.43 tend to have the highest temperature range, 
$T_\mathrm{warm} \sim 45\text{--}95$\,K. The values of $T_\mathrm{warm}$ 
are consistent with the fact that \hto\ traces warm gas. We did not find
any significant differences between the ranges of \nhto/$\Delta V$ derived 
from the modelling for these five sources, although G09v1.97 tends to have 
lower \nhto/$\Delta V$ (Table\,\ref{table:model_para}). As shown in Section\,\ref{AGN}, 
there is no evidence of AGN domination in all our sources, the submm \hto\ 
lines likely trace the warm dust component that connect to the heavily obscured 
active star-forming activity. However, due to the lack of photometry data on the 
Wien side of the dust SEDs, we will not be able to compare the observed values 
of $T_\mathrm{warm}$ directly with the ones derived from the modelling.

By adopting the 100\,$\mu$m dust mass absorption coefficient from 
\cite{2003ARA&A..41..241D} of $\kappa_{100}$\,=\,27.1\,cm$^2$\,g$^{-1}$,
we can derive the dust opacity by
\begin{equation} 
	\label{eq:tau}
       \tau_{100}=\kappa_{100} \, \sigma_\mathrm{dust}= \kappa_{100} 
       \left( M_\mathrm{dust} \over A \right) = \kappa_{100} 
       \left( M_\mathrm{dust} \over 2 \pi r_\mathrm{half}^2 \right)
\end{equation} 
where $\sigma_\mathrm{dust}$ is the dust mass column density, $M_\mathrm{dust}$ 
is the dust mass, $A$ is the projected surface area of the dust continuum 
source and $r_\mathrm{half}$ is the half-light radius of the source at submm. 
As shown in Table\,\ref{table:previous_obs_properties}, among the five sources 
in Fig.\,\ref{fig:h2o-model}, the values of $M_\mathrm{dust}$ and 
$r_\mathrm{half}$ in G09v1.97, NCv1.143 and NBv1.78 have been derived 
via gravitational lensing \citepalias{2013ApJ...779...25B}. Consequently, 
the derived approximate dust optical depth at 100\,$\mu$m in these three 
sources is $\tau_{100} \approx\,$\,1.8, 7.2 and 2.5, respectively. One should 
note that, the large uncertainty in both the $\kappa_{100}$ and $r_\mathrm{half}$ 
of these \hz\ galaxies can bring a factor of few error budget. Nevertheless, 
by adopting a gas-to-dust mass ratio of $X = 100$ 
\cite[e.g.][]{2011ApJ...740L..15M}, we can derive the gas depletion 
time using the following approach,
\begin{equation} 
	\label{eq:t_dep}
t_\mathrm{dep} = {M_\mathrm{gas} \over \mathit{SFR}} =     
                   {X \tau_{100} \over \Sigma_\mathit{SFR} \kappa_{100}}
                   \approx 1.8 \times 10^{4}  \left( \tau_{100} \over {\Sigma_\mathit{SFR} \over \mathrm{M_\odot\,yr^{-1}\,kpc^{-2}}} \right) \, \mathrm{Myr}                 
\end{equation} 
where $M_\mathrm{gas}$ is the total molecular gas mass and $\Sigma_\mathit{SFR}$ 
is the surface $\mathit{SFR}$ density derived from \lir\ using \citet{1998ARA&A..36..189K} 
calibration by assuming a Salpeter IMF 
\citepalias[][and Table\,\ref{table:previous_obs_properties}]{2013ApJ...779...25B}. 
The implied depletion time scale is $t_\mathrm{dep} \approx 35\text{--}60$\,Myr 
with errors within a factor of two, in which the dominant uncertainties are
from the assumed gas-to-dust mass ratio and the half-light radius. The $t_\mathrm{dep}$ 
is consistent with the values derived from dense gas tracers, like HCN in local 
(U)LIRGs \citep[e.g.][]{2004ApJ...606..271G, 2012A&A...539A...8G}. As suggested 
in \citetalias{2014A&A...567A..91G}, the \hto\ and HCN likely to be located in 
the same regions, indicate that the \hto\ traces the dense gas as well. Thus, 
the $\tau_{100}$ derived above is likely also tracing the \fir\ radiation source 
that powers the submm \hto\ emissions. \citetalias{2013ApJ...779...25B} also has 
found that these {\it H}-ATLAS \hz\ Hy/ULIRGs are expected to be optically thick
in the \fir. By adding the constrain from $\tau_{100}$ above, we can better derive 
the physical conditions in the sources as shown in Table\,\ref{table:model_para}.

\begin{table}[htbp]
\setlength{\tabcolsep}{0.64em}
\small
\caption{Parameters derived from \fir\ pumping model of \hto.}
\centering
\begin{tabular}{lrrrr}
\hline
\hline
Source   & $\tau_{100}$ & $T_\mathrm{warm}$ &     \nhto/$\Delta V$\;\;\;       & \nhto\;\;\;\;\;                  \\
         &              &       (K)\;\,     &     (cm$^{-2}$\,km$^{-1}$\,s)    & (cm$^{-2}$)\;\;\;\;              \\
\hline  
G09v1.97 &     1.8      &      45--55       &   (0.3--0.6)\,$\times 10^{15}$   & (0.3--1.1)\,$\times 10^{17}$     \\
G12v2.43 &     --       &      45--95       & $\gtrsim$\,0.7\,$\times 10^{15}$ & $\gtrsim$\,0.7\,$\times 10^{17}$ \\
NCv1.143 &     7.2      &      45--55       &   (2.0--20)\,$\times 10^{15}$    & (2.0--60)\,$\times 10^{17}$      \\
NAv1.177 &     --       &      45--75       & $\gtrsim$\,1.0\,$\times 10^{15}$ & $\gtrsim$\,1.0\,$\times 10^{17}$ \\
NBv1.78  &     2.5      &      45--75       & $\gtrsim$\,0.6\,$\times 10^{15}$ & $\gtrsim$\,0.6\,$\times 10^{17}$ \\
\hline  
\end{tabular}
\tablefoot{ $\tau_{100}$ is derived from Eq.\,\ref{eq:tau} with errors of a few units (see text), while 
			$T_\mathrm{warm}$ and \nhto/$\Delta V$ are inferred from the \hto\ excitation model. \nhto\
			is calculated by taking a typical $\Delta V$ value range of $100\text{--}300$\,km\,s$^{-1}$ 
			as suggested by \citetalias{2014A&A...567A..91G}.
       }
\label{table:model_para}
\end{table}
\normalsize

From their modelling of local infrared galaxies, \citetalias{2014A&A...567A..91G} 
find a range of $T_\mathrm{warm} =45\text{--}75$\,K, $\tau_{100}=0.05\text{--}0.2$ 
and \nhto/$\Delta V=(0.5\text{--}2)\times10^{15}$\,\,cm$^{-2}$\,km$^{-1}$\,s. The 
modelling results for our \hz\ sources are consistent with those in local galaxies 
in terms of $T_\mathrm{warm}$ and \nhto/$\Delta V$. However, the values of $\tau_{100}$ 
we found at \hz\ are higher than those of the local \ir\ galaxies. This is consistent 
with the higher ratio between \lhto\ and \lir\ at \hz\ \citepalias{2013ApJ...771L..24Y} 
which could be explained by higher $\tau_{100}$ \citepalias{2014A&A...567A..91G}. 
However, as demonstrated in an extreme sample, a very large velocity dispersion 
will also increase the value of \lhto/\lir\ within the sources with $\tau_{100} > 1$. 
Thus, the higher ratio can also be explained by larger velocity dispersion (not 
including systemic rotations) in the \hz\ Hy/ULIRGs. Compared with local ULIRGs, 
our {\it H}-ATLAS sources are much more powerful in terms of their \lir. The dense 
warm gas regions that \hto\ traces are highly obscured with much more powerful 
\fir\ radiation fields, which possibly are close to the limit of maximum starbursts. 
Given the values of dust temperature and dust opacity, the radiation pressure 
$P_\mathrm{rad} \sim \tau_{100} \sigma T_\mathrm{d} / c$ ($\sigma$ is 
Stefan-Boltzmann$'$s constant and $c$ the speed of light) of our sources is about 
$0.8 \times 10^{-7}$\,erg\,cm$^{-3}$. If we assume a H$_2$ density $n_\mathrm{H_2}$ 
of $\sim$\,10$^6$\,cm\,$^{-3}$ and take $T_\mathrm{k} \sim$\,150\,K as suggested 
in \citetalias{2014A&A...567A..91G}, the thermal pressure 
$P_\mathrm{th} \sim n_\mathrm{H_2} k_\mathrm{B} T_\mathrm{k} \sim 2 \times 10^{-8}$\,erg\,cm$^{-3}$
($k_\mathrm{B}$ is the Boltzmann constant and $T_\mathrm{k}$ is the gas temperature).
Assuming a turbulent velocity dispersion of 
$\sigma_\mathrm{v} \sim 20\text{--}50$\,km\,s$^{-1}$ \citep{2015A&A...575A..56B} 
and taking molecular gas mass density $\rho \sim 2 \mu n_\mathrm{H_2}$ (2$\mu$ 
is the average molecular mass) would yield for the turbulent pressure 
$P_\mathrm{turb} \sim \rho \sigma_\mathrm{v}^2/3 \sim 4 \times 10^{-6}$\,erg\,cm$^{-3}$.
This might be about an order of magnitude larger than $P_\mathrm{rad}$ and two 
orders of magnitude larger than $P_\mathrm{th}$, but we should note that all 
values are very uncertain, especially $P_\mathrm{turb}$ which could be 
uncertain by, at maximum, a factor of a few tens. Therefore, keeping in mind 
their large uncertainties, turbulence and/or radiation are likely to play an 
important role in limiting the star formation.

\subsection{Comparison between \hto\ and CO}
\label{CO lines}

The velocity-integrated flux density ratio between submm \hto\ and submm CO lines
with comparable frequencies is 0.02--0.03 in local PDRs such as Orion 
and M\,82 \citep{2010A&A...521L...1W}. But this ratio in local ULIRGs 
\citepalias{2013ApJ...771L..24Y} and in {\it H}-ATLAS \hz\ Hy/ULIRGs 
is much higher, from 0.4 to 1.1 (Table\,\ref{table:co_properties} and 
\ref{table:h2o_properties}). The former case is dominated by typical 
PDRs, where CO lines are much stronger than \hto\ lines, while the 
latter sources shows clearly a different excitation regime, in which 
\hto\ traces the central core of warm, dense and dusty molecular gas 
which is about a few hundred parsec \citep{2010A&A...518L..43G} in 
diameter in local ULIRGs and highly obscured even at \fir.

Generally, submm \hto\ lines are dominated by \fir\ pumping that 
traces strong \fir\ dust continuum emission, which is different from 
the regime of molecular gas traced by collisional excited CO lines. 
In the active star-forming nucleus of the \ir-bright galaxies, the 
\fir\ pumped \hto\ is expected to trace directly the \fir\ radiation 
generated by the intense star formation, which can be well correlated 
with the high-$J$ CO lines \citep{2015ApJ...810L..14L}. Thus there 
is likely to be a correlation between the submm \hto\ and CO emission. 
From our previous observations, most of the \hto\ and CO line profiles 
are quite similar from the same source in our \hz\ lensed Hy/ULIRGs 
sample (Fig.\,2 of \citetalias{2013A&A...551A.115O}). In the present 
work, we again find similar profiles between \hto\ and CO in terms 
of their FWHM with an extended sample (Table\,\ref{table:co_properties} 
and \ref{table:h2o_properties}). In both cases the FWHMs of \hto\ and 
CO are generally equal within typical 1.5\,$\sigma$ errors (see special 
discussion for each source in Appendix\,\ref{Individual sources}).

As the gravitational lensing magnification factor is sensitive 
to spatial alignment, the similar line profiles could thus suggest 
similar spatial distributions of the two gas tracers. However, 
there are a few exceptional sources, such as SDP\,81 \citep{2015ApJ...808L...4A} 
and HLSJ0918 \citep{2014ApJ...783...59R}. In both cases, the \hto\ 
lines are lacking the blue velocity component found in the CO line 
profiles. Quite different from the rest sources, in SDP\,81 and HLSJ0918, 
the CO line profiles are complicated with multiple velocity components. 
Moreover, the velocity-integrated flux density ratios between these CO components may vary 
following the excitation level (different $J_\mathrm{up}$). Thus, 
it is important to analyse the relation between different CO excitation 
components (from low-$J$ to high-$J$) and \hto. Also, high resolution 
observation is needed to resolve the multiple spatial gas components 
and compare the CO emission with \hto\ and dust continuum emission 
within each component. 



\subsection{AGN content}
\label{AGN}
It is still not clear how a strong AGN could affect the excitation of 
submm \hto\ in both local ULIRGs and \hz\ Hy/ULIRGs. Nevertheless, there 
are some individual studies addressing this question. For example, in 
APM\,08279, \cite{2011ApJ...741L..38V} found that AGN is the main power 
source that excites the high-$J$ \hto\ lines and also enriches the 
gas-phase \hto\ abundance. Similar conclusion has also been drawn by 
\cite{2010A&A...518L..43G} that in Mrk\,231 the AGN accounts for at 
least 50\,\% contribution to the \fir\ radiation that excites \hto. 
From the systematic study of local sources \citepalias{2013ApJ...771L..24Y}, 
slightly lower values of \lhto/\lir\ are found in strong-AGN-dominated 
sources. In the present work, the decreasing ratio of \lhto/\lir\ with AGN is 
clearly shown in Fig.\,\ref{fig:h2o-ir} where Mrk\,231, G09v1.124-W 
and APM\,08279 are below the correlation by factors between 2 and 5 with
less than 30\% uncertainties (except the \htot321123 line of Mrk\,231).

In the \fir\ pumping regime, the buried AGN will provide a strong \fir\ 
radiation source that will pump the \hto\ lines. However, the very warm 
dust powered by the AGN will increase the value of \lir\ faster than the 
number of $\geq$\,75\,$\mu m$ photons that is dominating the excitation 
of $J \leq 3$ \hto\ lines \citep[e.g.][]{2015ApJ...814....9K}. If we assume 
that the strength of the \hto\ emission is proportional to the number of 
pumping photons, then in the strong-AGN-dominated sources, the ratio of 
\lhto/\lir\ will decrease since much warmer dust is present. Moreover, strong 
radiation from the AGN could dissociate the \hto\ molecules.

To evaluate the AGN contribution to the {\it H}-ATLAS sources, we extracted 
the 1.4\,GHz radio flux from the FIRST radio survey \citep{1995ApJ...450..559B} 
listed in Table\,\ref{table:previous_obs_properties}. By comparing the 
\fir\ and radio emission using the q parameter 
\citep{1992ARA&A..30..575C},
$q \equiv \log ( {L_\mathrm{FIR} / 3.75 \times 10^{12}\,\mathrm{W}} ) 
 - \log ( {{L_\mathrm{1.4\,GHz}} / {1\,\mathrm{W}\,\mathrm{Hz}^{-1}}} )$,
we derive values of $q$ from 1.9 to 2.5 in our sources. These values 
follow the value $2.3\pm0.1$ found by \citet{2001ApJ...554..803Y} for non 
strong-radio AGN. This may suggest that there is also no significant indication 
of a high radio contribution from AGN. This is also confirmed by the 
Wide-field Infrared Survey Explorer \citep[WISE,][]{2010AJ....140.1868W}, 
which does not detect our sources at 12\,$\mu$m and 22\,$\mu$m. 
However, rest-frame optical spectral observations show that 
G09v1.124-W is rather a powerful AGN (Oteo et al, in prep.), 
which is the only identified AGN-dominated source in our sample.

\section{Detection of \hto$^+$ emission lines}
\label{htop}

 \begin{figure*}[htbp]
	 \begin{center}
 \includegraphics[scale=0.86]{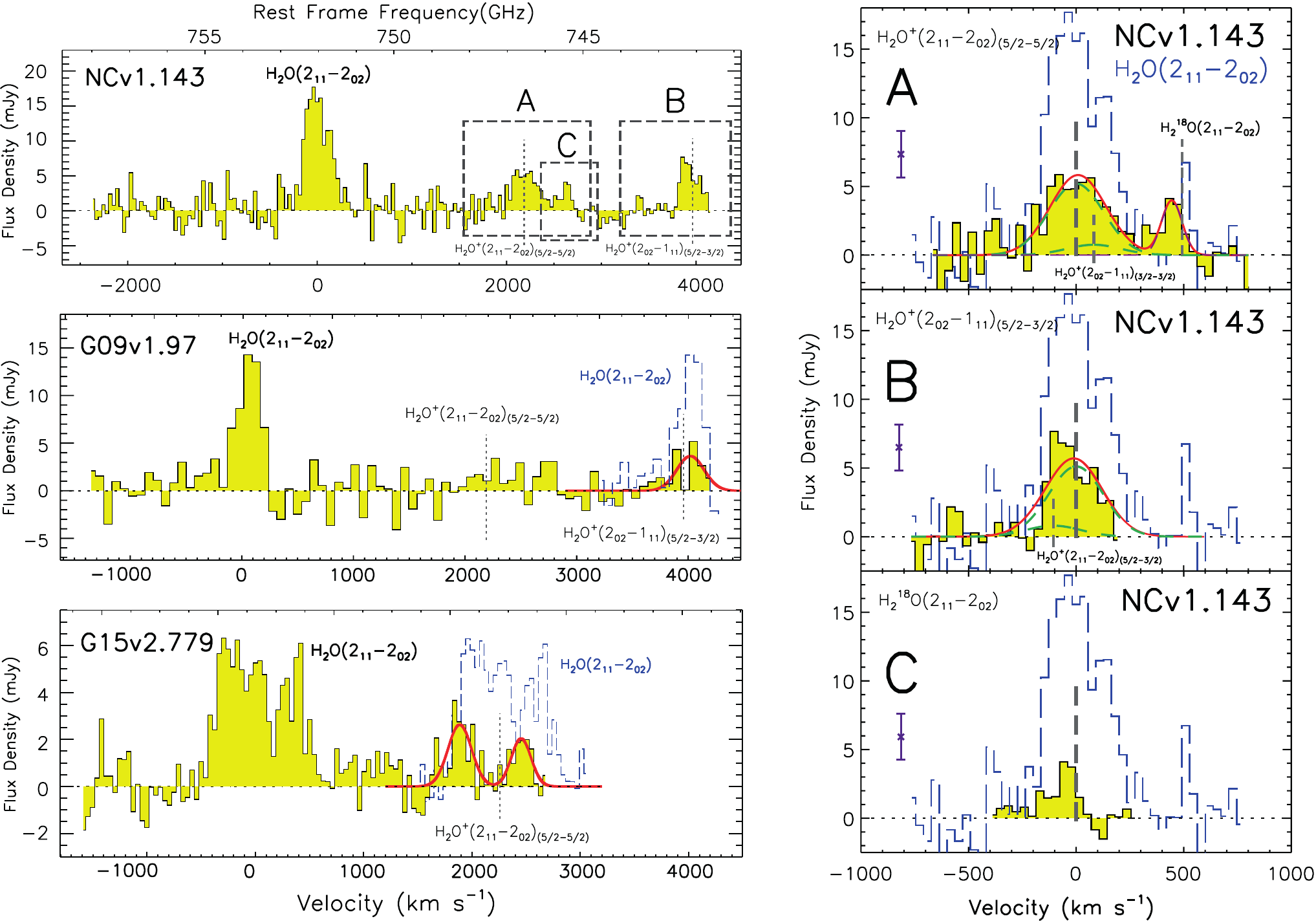}
 \caption{
         {\em Left panel}: from top to bottom are the full NOEMA 
         spectrum at $\nu_\mathrm{rest} \sim 750$\,GHz of NCv1.143, 
         G09v1.97 and G15v2.779, respectively. The reference frequency 
         is the redshifted frequency of the line \htot211202. The 
         frequencies of the main \htop(\t211202)\,$_{(5/2-5/2)}$ and 
         \htop(\t202111)\,$_{(5/2-3/2)}$ lines are indicated by 
         grey vertical dashed lines. The three dashed squares in 
         the spectrum of NCv1.143 show the position of each zoom-in
         spectrum of the \htop\ (or the H$_2^{18}$O) as displayed 
         in the {\it right panel} indicated by the A, B or C. The superposed 
         blue dashed histograms represents the spectra of \htot211202 
         centred at the frequencies of the \htop\ lines. Note that, 
         in many cases, the observed frequency ranges (yellow histograms) 
         do not include the full expected profiles for the \htop\ lines. 
         The red curve represents the Gaussian fitting to the spectra. 
         We have detected both \htop\ lines in NCv1.143, and
         tentatively detected \htop(\t202111)\,$_{(5/2-3/2)}$ in G09v1.97
         and \htop(\t211202)\,$_{(5/2-5/2)}$ in G15v2.779.
         {\em Right panel}: from top to bottom are the spectra dominated by 
          lines of \htop(\t211202)\,${_{(5/2-5/2)}}$, \htop(\t202111)\,$_{(3/2-3/2)}$ 
          and H$_2^{18}$O(\t211202), respectively, displayed as the 
          filled yellow histograms. The reference frequency is the frequency 
          of each of these lines. Weaker \htop(\t202111)\,$_{(3/2-3/2)}$ and 
		  \htop(\t211202)\,$_{(5/2-3/2)}$ components are indicated by additional
		  grey vertical dashed lines. The superposed blue dashed histograms 
		  represent the spectra of para-\htot211202 in NCv1.143 centred at each 
		  line frequency. The red curve represents the Gaussian fitting 
		  to the spectra, and the green dashed curves are the decomposed 
		  Gaussian profiles for each fine structure line. The violet error 
		  bar indicates the $\pm$\,1\,$\sigma$ uncertainties of the spectrum.
         }
 \label{fig:h2o-ions1}
     \end{center}
 \end{figure*}

H$_2$O can be formed through both solid-state and gas-phase chemical 
reactions \citep{2013ChRv..113.9043V}. On dust-grain mantles, surface 
chemistry dominates the formation of \hto\ molecules. Then they can be 
released into the interstellar medium (ISM) gas through sublimation. 
In the gas phase, \hto\ can be produced through two routes: the 
neutral-neutral reaction, 
usually related to shocks, creates \hto\ via \ce{O + H2 -> OH + H}; 
\ce{OH + H2 -> H2O + H} at high temperature ($\gtrsim$\,300\,K). At 
lower temperature ($\lesssim$\,100\,K), the ion-neutral reactions in 
photon-dominated regions (PDRs), cosmic-ray-dominated regions and 
X-ray-dominated regions \citep[e.g.][]{2005A&A...436..397M} generate 
\hto\ from O, H$^+$, H$_3^+$ and H$_2$, with intermediates such as 
O$^+$, OH$^+$, \htop\ and H$_3$O$^+$, and finally \ce {H3O^+ + e -> H2O + H}. 
However, classical PDRs are not likely linked to these highly excited 
submm \hto\ emissions \citepalias{2013ApJ...771L..24Y}. Therefore, 
\htop\ lines are important for distinguishing between shock- or 
ion-chemistry origin for \hto\ in the early Universe, indicating 
the type of physical regions in these galaxies: shock-dominated 
regions, cosmic-ray-dominated regions or X-ray-dominated regions. 
Indeed, they can be among the most direct tracers of the cosmic-ray 
or/and X-ray ionization rate 
\citep[e.g.][]{2010A&A...518L.110G, 2010A&A...521L..10N, 2013A&A...550A..25G} 
of the ISM, which regulates the chemistry and influences many key 
parameters, for example, X-factor \citep{2007MNRAS.378..983B} that connects 
the CO luminosity to the H$_2$ mass. Moreover, the significant detections 
of \htop\ emission in \hz\ Hy/ULIRGs could help us understanding \hto\ 
formation in the early Universe.

When observing our sources with redshift $z \gtrsim 3.3$, it is possible 
to cover all the following lines with the NOEMA WideX bandwidth: para-\htot211202 
at 752\,GHz and four ortho-\htop\ lines (two intertwined fine structure doublets 
of two different lines whose frequencies almost coincide by chance): 
\t202111\,$_{(5/2-3/2)}$ at 742.1\,GHz, \t211202\,$_{(5/2-3/2)}$ at 742.3\,GHz, 
\t202111\,$_{(3/2-3/2)}$ at 746.3\,GHz and \t211202\,$_{(5/2-5/2)}$ at 746.5\,GHz, 
in the 3.6\,GHz band simultaneously (the rest-frame frequencies are taken from the 
CDMS catalogue: \url{http://www.astro.uni-koeln.de/cdms}, see energy level 
diagram of \htop\ in Fig.\,\ref{fig:h2o-e-level} and the 
full spectra in Fig.\,\ref{fig:h2o-ions1}). 
Additionally, within this range, we can also cover the H$_2^{18}$O(\t211202) 
line at 745.3\,GHz. There are three sources of our sample
that have been observed in such a frequency setup: NCv1.143, NCv1.268 and 
G09v1.97. We have also included the source G15v2.779 from our previous 
observation \citepalias{2013A&A...551A.115O}, in which we have covered both 
\htot211202 at 752\,GHz and \htop\ lines around 746\,GHz. 
We have detected both main lines of \htop\ in NCv1.143, and tentatively 
detected one line in G09v1.97 and G15v2.779 (Fig.\,\ref{fig:h2o-ions1}).
For NCv1.268, due to the large noise level and the complex line 
profile, we were not able to really identify any \htop\ line detection.

\begin{table}[htbp]
\setlength{\tabcolsep}{0.65em}
\small
\caption{Observed ortho-\htop\ fine structure line parameters of the \hz\ {\it H}-ATLAS lensed HyLIRGs.}
\centering
\begin{tabular}{lcrrr}
\hline
\hline
Source             &  \htop\ transition        & $\nu_{\rm rest}$ & $\nu_{\rm line}$  & $I_{\rm H_{2}O^+}$ \\
                   &                           &    (GHz)         &    (GHz)          & (Jy\,km\,s$^{-1}$) \\
\hline  
NCv1.143           &  \t211202\,$_{(5/2-5/2)}$ &   746.5          &   163.53          & $1.6\pm0.5$        \\
                   &  \t202111\,$_{(3/2-3/2)}$ &   746.3          &   163.48          & $0.2\pm0.5$        \\
				   &  \t211202\,$_{(5/2-3/2)}$ &   742.3          &   162.61          & $0.3\pm0.4$        \\
				   &  \t202111\,$_{(5/2-3/2)}$ &   742.1          &   162.56          & $1.6\pm0.4$        \\
G09v1.97           &  \t202111\,$_{(5/2-3/2)}$ &   742.1          &   160.14          & $1.4\pm0.4$        \\
G15v2.779          &  \t211202\,$_{(5/2-5/2)}$ &   746.5          &   142.35          & $1.2\pm0.3$        \\
\hline  
\end{tabular}
\tablefoot{ 
	    The \htop\,(\t202111)\,$_{(5/2-3/2)}$ line in G09v1.97 is 
	    blended by (\t211202)\,$_{(5/2-3/2)}$, and 
	    \htop\,(\t211202)\,$_{(5/2-5/2)}$ line in G15v2.779 is 
	    blended by (\t202111)\,$_{(3/2-3/2)}$. However, the contribution 
	    from the latter in each case is small, likely less than 
	    20\,\% as shown in the case of the \htop\ lines in NCv1.143.
	    Note that the quoted uncertainties do not include the missing 
	    parts of the spectra cut by the limited observed bandwidth 
	    (Fig.\,\ref{fig:h2o-ions1}).
       }
\label{table:htop_intensity}
\end{table}
\normalsize

As shown in Fig.\,\ref{fig:h2o-ions1}, in NCv1.143, the dominant 
\htop\ fine structure lines \t211202\,$_{(5/2-5/2)}$ at 746.5\,GHz 
and \t202111\,$_{(5/2-3/2)}$ at 742.1\,GHz are well detected. 
The velocity-integrated flux densities of the two lines from a 
two-Gaussian fit are $1.9\pm0.3$ and $1.6\pm0.2$\,Jy\,km\,s$^{-1}$, 
respectively. These are the approximate 
velocity-integrated flux densities of the dominant \htop\ lines \t211202\,$_{(5/2-5/2)}$ 
and \t202111\,$_{(5/2-3/2)}$ if neglecting the minor 
contributions from \htop\ lines \t202111\,$_{(3/2-3/2)}$ at 746.2\,GHz 
and \t211202\,$_{(5/2-3/2)}$ at 742.3\,GHz. However, the \htop\ line 
profile at 746.5\,GHz is slightly wider than the \hto\ line 
(Fig.\,\ref{fig:h2o-ions1}), probably due to a contribution from the 
fairly weak fine structure line \htop(\t202111)\,$_{(3/2-3/2)}$ at 
746.3\,GHz. The ratio between total velocity-integrated flux density of the \htop\ 
lines and \htot211202 is $0.60\pm0.07$ (roughly 0.3 for each dominant 
\htop\ line), being consistent with the average value from the local 
\ir\ galaxies \citepalias{2013ApJ...771L..24Y}\footnote{As suggested by 
\cite{2013A&A...550A..25G}, due to the very limited spectral resolution 
of {\it Herschel}/SPIRE FTS, the ortho-\htop(\t202111)\,$_{(3/2-3/2)}$ 
line at 746.5\,GHz quoted in \citetalias{2013ApJ...771L..24Y} is 
actually dominated by ortho-\htop(\t211202)\,$_{(5/2-5/2)}$, 
considering their likely excitation and relative strength.}. In order
to derive the velocity-integrated flux density of each fine structure doublets around 742 
and 746\,GHz, we have also performed a four-Gaussian fit with fixed 
line positions (equal to $\nu_\mathrm{rest}/(1+z)$) and linewidth 
(equals to that of \htot211202). We find the velocity-integrated flux densities
of the two fine structure lines of \htop(\t211202) are $1.6\pm0.5$ 
and $0.3\pm0.4$\,Jy/km\,s$^{-1}$, while they are $1.6\pm0.4$ and 
$0.2\pm0.5$\,Jy/km\,s$^{-1}$ for the two fine structure lines of 
\htop(\t202111) (Table\,\ref{table:htop_intensity}).
We should note that these fitting results have much larger 
uncertainties due to the blending. Nevertheless, they are consistent 
with the earlier fitting results without de-blending. The similarity 
of the velocity-integrated flux densities between the \htop(\t202111) and \htop(\t211202) 
lines is in good agreement with the regime of \fir\ pumping 
as submm \hto\ \citep{2013A&A...550A..25G}. As a first approximation, 
if these \htop\ lines are optically thin and we ignore the additional 
pumping from ortho-\htop\ 2$_{02}$ to ortho-\htop\ $J=3$ energy levels, 
the statistical equilibrium applied to energy level 2$_{02\,5/2}$ 
implies that all population arriving per second at 2$_{02\,5/2}$ 
should be equal to all population leaving the level per second. 

After subtracting the Gaussian profiles of all the \htop\ lines
in the spectrum, we find a $3$\,$\sigma$ residual in terms
of the velocity-integrated flux density around 745.3\,GHz 
($I = 0.6\pm0.2$\,Jy\,km\,s$^{-1}$, see Fig.\ref{fig:h2o-ions1}). 
This could be a tentative detection of the H$_2^{18}$O(\t211202) 
line at 745.320\,GHz. The velocity-integrated flux density ratio 
of H$_2^{18}$O(\t211202) over \htot211202 in NCv1.143 would hence 
be $\sim0.1$. If this tentative detection was confirmed, it would 
show that ALMA could easily study such lines. But sophisticated 
models will be needed to infer isotope ratios.

\begin{figure}[htbp]
	 \begin{center}
 \includegraphics[scale=0.441]{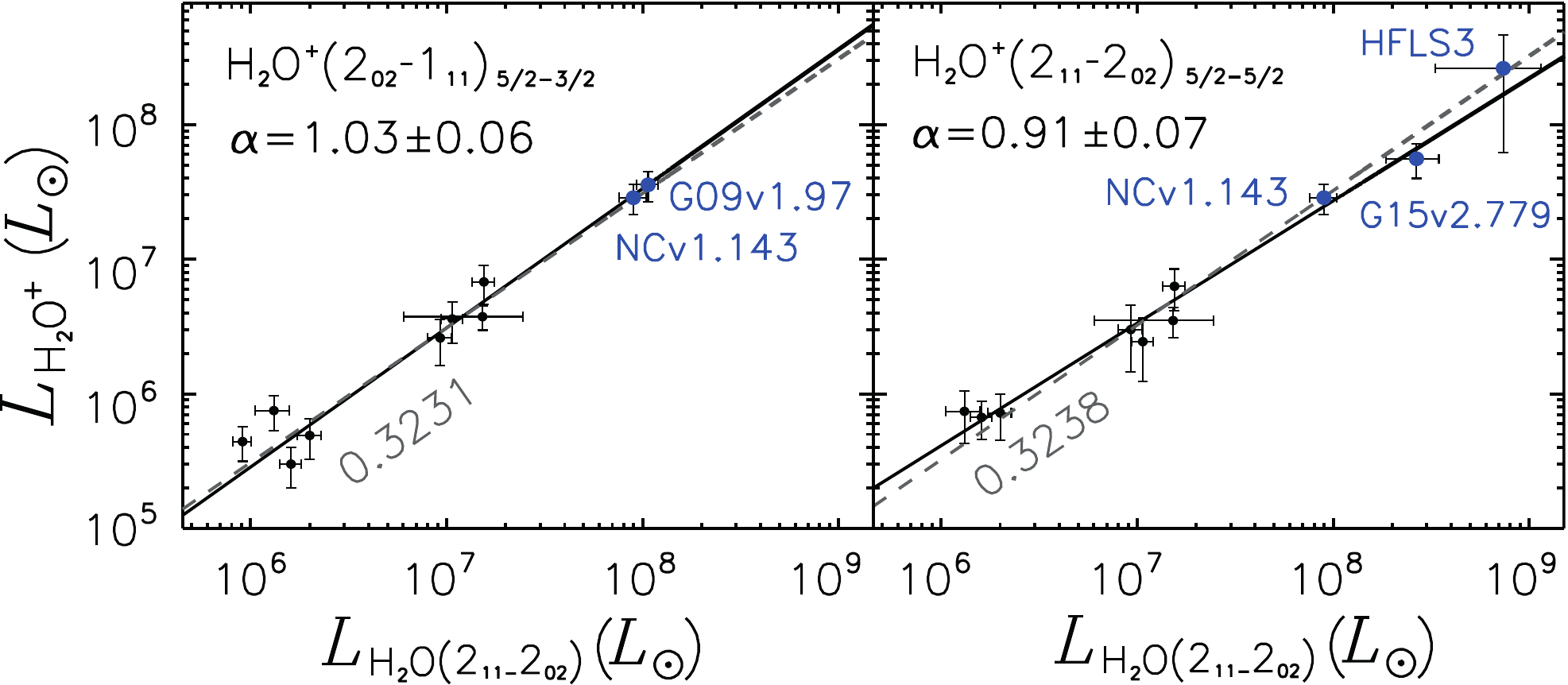}
 \caption{
		  Correlation between the luminosity of $J=2$ ortho-\htop\ 
		  and para-\htot211202. The fitted function is 
		  \lhtop\;$\propto$\;\lhto$^{\alpha}$. We found a very good 
		  correlation between \lhtop\ and \lhto\ with a slope close
		  to one. Black points are from the local 
		  ULIRGs as listed in Table\,\ref{table:htop}. Dark blue 
		  ones are \hz\ starbursts from this work. Black solid 
		  lines indicate the $\chi^2$ fitting results while the 
		  grey dashed lines and the grey annotations represent 
		  the average ratio between the \lhtop\ and \lhto.   
         }
 \label{fig:h2op-h2o}
     \end{center}
 \end{figure}

The spectrum of the \htot211202 line in G09v1.97 covers both 
the two main \htop\ fine structure lines (Fig\,\ref{fig:h2o-ions1}). 
However, due to the 
limited sensitivity, we have only tentatively detected the
\htop(\t202111)\,$_{(5/2-3/2)}$ line just above 3\,$\sigma$ 
(neglecting the minor contribution from \htop(\t211202)\,$_{(5/2-3/2)}$), 
and the velocity-integrated flux density is $1.4\pm0.4$\,Jy\,km\,s$^{-1}$ using a
single Gaussian fit. We did not perform any line de-blending for 
this source considering the data quality. The \htop\ line 
profile is in good agreement with that of the \hto\ (blue dashed 
histogram in Fig.\,\ref{fig:h2op-h2o}). The velocity-integrated flux density 
of the undetected \htop(\t211202)\,$_{(5/2-5/2)}$ line could also be 
close to this value as discussed in the case of NCv1.143, yet somewhat lower 
and not detected in this source. More sensitive observation should 
be conducted to further derive robust line parameters.

We have also tentatively detected the \htop(\t211202)\,$_{(5/2-5/2)}$ 
line in G15v2.779 ($S/N\;\sim 4$ by neglecting the minor contribution 
from the \htop(\t202111)\,$_{(3/2-3/2)}$ line). The line 
profile is in good agreement with that of \htot211202 (blue dashed 
histogram in Fig.\,\ref{fig:h2o-ions1}). The velocity-integrated flux density 
derived from a double-peak Gaussian fit is $1.2\pm0.3$\,Jy\,km\,s$^{-1}$ (we did 
not perform any line de-blending for the \htop\ doublet considering 
the spectral noise level). There could be a minor contribution from 
the \htop(\t202111)\,$_{(3/2-3/2)}$ line to the velocity-integrated 
flux density. However, such a contribution is likely to be negligible 
as in the case of NCv1.143. The contribution is also within the 
uncertainty of the velocity-integrated flux density. Nevertheless, 
the position of \htop\ has a small blue-shift compared with \hto, 
but note that the blue part of the line is cut by the limited 
observed bandwidth (yellow histogram).

After including the local detections of \htop\ lines from 
\citetalias{2013ApJ...771L..24Y} (Table\,\ref{table:htop}),
we find a tight linear correlation between the luminosity of \hto\ 
and the two main lines of \htop\ (slopes equal to $1.03\pm0.06$ and 
$0.91\pm0.07$, see Fig.\,\ref{fig:h2op-h2o}). However, one should keep 
in mind that, because the local measurement done by {\it Herschel} 
SPIRE/FTS \citep{2010SPIE.7731E..16N} has rather low spectral resolution, 
neither \htop(\t211202)\,$_{(5/2-3/2)}$ and \htop(\t202111)\,$_{(5/2-3/2)}$, 
nor \htop(\t211202)\,$_{(5/2-5/2)}$ and \htop(\t202111)\,$_{(3/2-3/2)}$ can be 
spectroscopically resolved. In the correlation plot (Fig.\,\ref{fig:h2op-h2o}) 
and Table\,\ref{table:htop}, we use the total luminosity from the 742\,GHz and 
746\,GHz lines, by assuming the contribution from \htop(\t211202)\,$_{(5/2-3/2)}$ 
and \htop(\t202111)\,$_{(3/2-3/2)}$ to the velocity-integrated flux density 
of the line at 742\,GHz and 746\,GHz is small ($\sim18$\,\%) and does not 
vary significantly between different sources. Hence, the velocity-integrated 
flux density ratio between each of the two dominant \htop\ fine structure 
lines and \hto\ in NCv1.143, G15v2.779 and G09v1.97 is $\sim 0.3$ 
(uncertainties are less than 30\%), which is consistent with local 
galaxies as shown in the figure. This ratio is much larger than the abundance 
ratio of \htop/\hto\,$\sim$\,0.05 found in Arp\,220, an analogue of \hz\ 
ULIRGs \citep{2011ApJ...743...94R}.

As discussed above, the AGN contribution to the excitation of the submm 
lines of most of our sources appears to be minor. Thus, the formation 
of \htop\ is likely dominated by cosmic-ray ionization, rather than 
X-ray ionization. Given the average luminosity ratio of 
\htop/\hto\;$\sim 0.3\pm0.1$ shown in Fig.\,\ref{fig:h2op-h2o}, \citet{2011A&A...525A.119M} 
suggest a cosmic-ray ionization rate of 10$^{-14}$--10$^{-13}$\,s$^{-1}$. 
Such high cosmic-ray ionization rates drive the ambient ionization 
degree of the ISM to 10$^{-3}$--10$^{-2}$, rather than the canonical 
10$^{-4}$. Therefore, in the gas phase, an ion-neutral route likely 
dominates the formation of \hto. However, \hto\ can also be enriched 
through the water-ice sublimation that releases \hto\ into the gas-phase 
ISM. As the upper part, $\sim 90$\,K, of the possible range for 
$T_\mathrm{warm}$ is close to the sublimation 
temperature of water ice. Hence, the high \hto\ abundance 
(\nhto\;$\gtrapprox 0.3 \times 10^{17}$\,cm$^{-2}$, see Section\,\ref{hto}) 
observed is likely to be the result of ion chemistry dominated by 
high cosmic-ray ionization and/or perhaps water ice desorption. However, 
further observation of \htop\ lines of different transitions and a 
larger sample is needed to constrain the contribution to \hto\ 
formation from neutral-neutral reactions dominated by shocks.

\section{Conclusions}
\normalsize
\label{Conclusions}
In this paper, we report a survey of submm \hto\ emission at redshift 
$z \sim 2\text{--}4$, by observing a higher excited ortho-\htot321312 
in 6 sources and several complementary $J=2$ para-\hto\ emission lines 
in the warm dense cores of 11 \hz\ lensed extreme starburst galaxies 
(Hy/ULIRGs) discovered by {\it H}-ATLAS. So far, we have detected an 
\hto\ line in most of our observations of a total sample of 17 \hz\ 
lensed galaxies, in other words, we have detected both $J=2$ para-\hto\ and $J=3$ 
ortho-\hto\ lines in five, and in ten other sources only one $J=2$ 
para-\hto\ line. In these \hz\ Hy/ULIRGs, \hto\ is the second strongest 
molecular emitter after CO within the submm band, as in local ULIRGs. 
The spatially integrated \hto\ emission lines have a velocity-integrated 
flux density ranging from 4 to 15\,Jy\,km\,s$^{-1}$, which yields the 
apparent \hto\ emission luminosity, $\mu$\lhto\, ranging from 
$\sim 6\text{--}22 \times 10^{8}$\,\lsun. After correction for gravitation 
lensing magnification, we obtained the intrinsic \lhto\ for para-\hto\ 
lines \t202111, \t211202 and ortho-\htot321312. The luminosities of the 
three \hto\ lines increase with \lir\ as 
\lhto\;$\propto$\;\lir$^{1.1\text{--}1.2}$. This correlation indicates 
the importance of \fir\ pumping as a dominant mechanism of \hto\ 
excitation. Comparing with $J=3$ to $J=6$ CO lines, the linewidths 
between \hto\ and CO are similar, and the velocity-integrated flux 
densities of \hto\ and CO are comparable. The similarity in line profiles 
suggests that these two molecular species possibly trace similar intense 
star-forming regions.

Using the \fir\ pumping model, we have analysed the ratios between $J=2$ and 
$J=3$ \hto\ lines and \lhto/\lir\ in 5 sources with both $J$ \hto\ lines detected. 
We have derived the ranges of the warm dust temperature ($T_\mathrm{warm}$), the \hto\ 
column density per unit velocity interval (\nhto/$\Delta V$) and the optical depth 
at 100\,$\mu$m ($\tau_{100}$). Although there are strong degeneracies, these 
modelling efforts confirm that, similar to 
those of local ULIRGs, these submm \hto\ emissions in \hz\ Hy/ULIRGs trace the 
warm dense gas that is tightly correlated with the massive star forming activity. 
While the values of $T_\mathrm{warm}$ and \nhto\ (by assuming that they have 
similar velocity dispersion $\Delta V$) are similar to the local ones, $\tau_{100}$ 
in the \hz\ Hy/ULIRGs is likely to be greater than 1 (optically thick), which 
is larger than $\tau_{100}=0.05\text{--}0.2$ found in the local \ir\ galaxies. 
However, we notice that the parameter space is still not well constrained in 
our sources through \hto\ excitation modelling. Due to the limited excitation 
levels of the detected \hto\ lines, we are only able to perform the modelling 
with pure \fir\ pumping.

The detection of relatively strong \htop\ lines opens the possibility to 
help understanding the formation of such large amount of \hto. In these 
\hz\ Hy/ULIRGs, the \hto\ formation is likely to be dominated by ion-neutral 
reactions powered by cosmic-ray-dominated regions. 
The velocity-integrated flux density ratio between \htop\ and \hto\ 
($I_{\mathrm{H_2O^+}}/\ihto \sim 0.3$), is remarkably constant from low to 
high-redshift, reflecting similar conditions in Hy/ULIRGs. However, 
more observations of \htop\ emission/absorption and also OH$^{+}$ 
lines are needed to further constrain the physical parameters of the 
cosmic-ray-dominated regions and the ionization rate in those regions.

We have demonstrated that the submm \hto\ emission lines are strong and 
easily detectable with NOEMA. Being a unique diagnostic, the \hto\ 
emission offers us a new approach to constrain the physical conditions 
in the intense and heavily obscured star-forming regions dominated by 
\fir\ radiation at \hz. Follow-up observations of other gas tracers, 
for instance, CO, HCN, \htop\ and OH$^+$ using the NOEMA, IRAM 30m and JVLA 
will complement the \hto\ diagnostic of the structure of different 
components, dominant physical processes, star formation and chemistry 
in \hz\ Hy/ULIRGs. 

With unprecedented spatial resolution and sensitivity, the image from 
the ALMA long baseline campaign observation of SDP\,81 \citep[also known 
as {\it H}-ATLAS\,J090311.6+003906, ][]{2015ApJ...808L...4A, 2015MNRAS.452.2258D,
2015MNRAS.451L..40R}, shows the resolved structure of the dust, CO and 
\hto\ emission in the $z=3$ ULIRG. With careful reconstruction of the 
source plane images, ALMA will help to resolve the submm \hto\ emission 
in \hz\ galaxies into the scale of giant molecular clouds, and 
provide a fresh view of detailed physics and chemistry in the early 
Universe.

\begin{acknowledgement}
We thank our referee for the very detail comments and suggestions 
which have improved the paper. This work was based on observations 
carried out with the IRAM Interferometer 
NOEMA, supported by INSU/CNRS (France), MPG (Germany), and IGN (Spain). 
The authors are grateful to the IRAM staff for their support. CY thanks 
Claudia Marka and Nicolas Billot for their help of the IRAM 30m/EMIR 
observation. CY also thanks Zhi-Yu Zhang and Iv\'an Oteo for insightful 
discussions. CY, AO and YG acknowledge support by NSFC grants \#11311130491,
\#11420101002 and CAS Pilot B program \#XDB09000000. CY and YG also 
acknowledge support by NSFC grants \#11173059. CY, AO, AB 
and YG acknowledge support from the Sino-French LIA-Origin joint exchange 
program. E.G-A is a Research Associate at the Harvard-Smithsonian
Center for Astrophysics, and thanks the Spanish Ministerio de 
Econom\'{\i}a y Competitividad for support under projects
FIS2012-39162-C06-01 and  ESP2015-65597-C4-1-R, and NASA grant ADAP
NNX15AE56G. RJI acknowledges support from ERC in the form of the Advanced 
Investigator Programme, 321302, COSMICISM. US participants in {\it H}-ATLAS 
acknowledge support from NASA through a contract from JPL. Italian 
participants in {\it H}-ATLAS acknowledge a financial contribution 
from the agreement ASI-INAF I/009/10/0. SPIRE has been developed by 
a consortium of institutes led by Cardiff Univ. (UK) and including: 
Univ. Lethbridge (Canada); NAOC (China); CEA, LAM (France); IFSI, 
Univ. Padua (Italy); IAC (Spain); Stockholm Observatory (Sweden); 
Imperial College London, RAL, UCL-MSSL, UKATC, Univ. Sussex (UK); 
and Caltech, JPL, NHSC, Univ. Colorado (USA). This development has 
been supported by national funding agencies: CSA (Canada); NAOC (China); 
CEA, CNES, CNRS (France); ASI (Italy); MCINN (Spain); SNSB (Sweden); 
STFC, UKSA (UK); and NASA (USA). CY is supported by the China 
Scholarship Council grant (CSC No.201404910443). 
\end{acknowledgement}

\footnotesize
\bibliographystyle{aa}
\bibliography{ms} 
\normalsize

\begin{appendix}


\section{Individual sources}

\begin{subfigures}
 \begin{figure*}[htbp]
	 \begin{center}
 \includegraphics[scale=1.0]{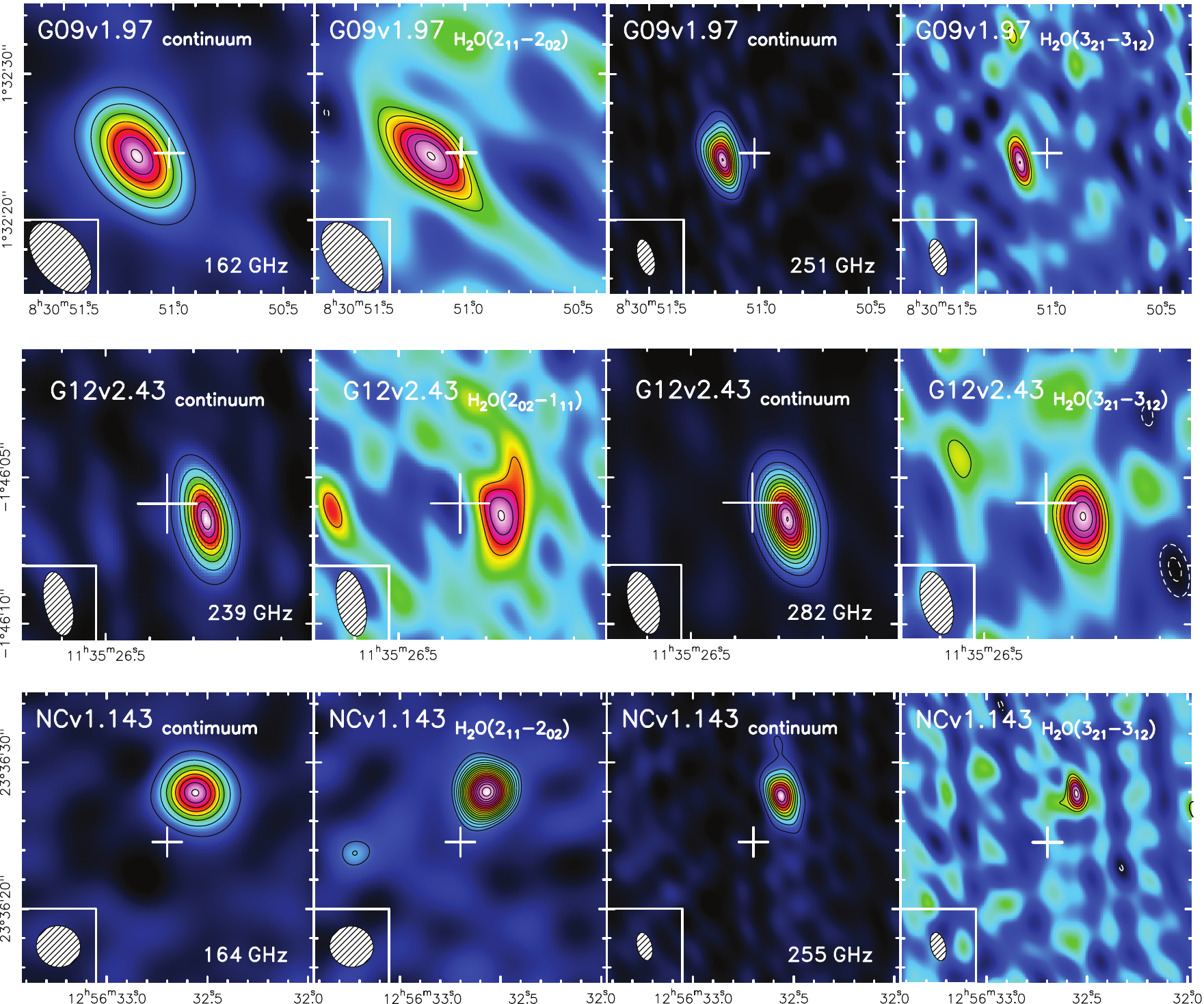}
 \caption{
		  Mapping of the \hto\ emission lines and the corresponding continuum emission 
		  (frequencies have been shown accordingly in the white text) 
		  in the sources with both para $J=2$ and ortho $J=3$ \hto\ lines observed. 
		  The contours of the continuum emission start from $6\,\sigma$ in step of 
		  $10\,\sigma$, and the contours of the \hto\ emission start from $3\,\sigma$ 
		  in step of $1\,\sigma$. Asymmetric negative contours are shown in white dashed 
		  lines. For each observation, the $1\,\sigma$ contours for 
		  the continuum (mJy\,beam$^{-1}$) and the \hto\ emission line 
		  (Jy\,km\,s$^{-1}$\,beam$^{-1}$) are as follows: G09v1.97 \htot211202 
		  (0.17/0.57), G09v1.97 \htot321312 (0.25/0.38), G12v2.43 \htot202111 (0.29/0.48), 
		  G12v2.43 \htot321312 (0.30/0.53), NCv1.143 \htot211202 (0.16/0.36) and
		  NCv1.143 \htot321312 (0.42/0.72).
         }
  \label{fig:map-1}
     \end{center}
 \end{figure*}

 \begin{figure*}[htbp]
	 \begin{center}
 \includegraphics[scale=1.0]{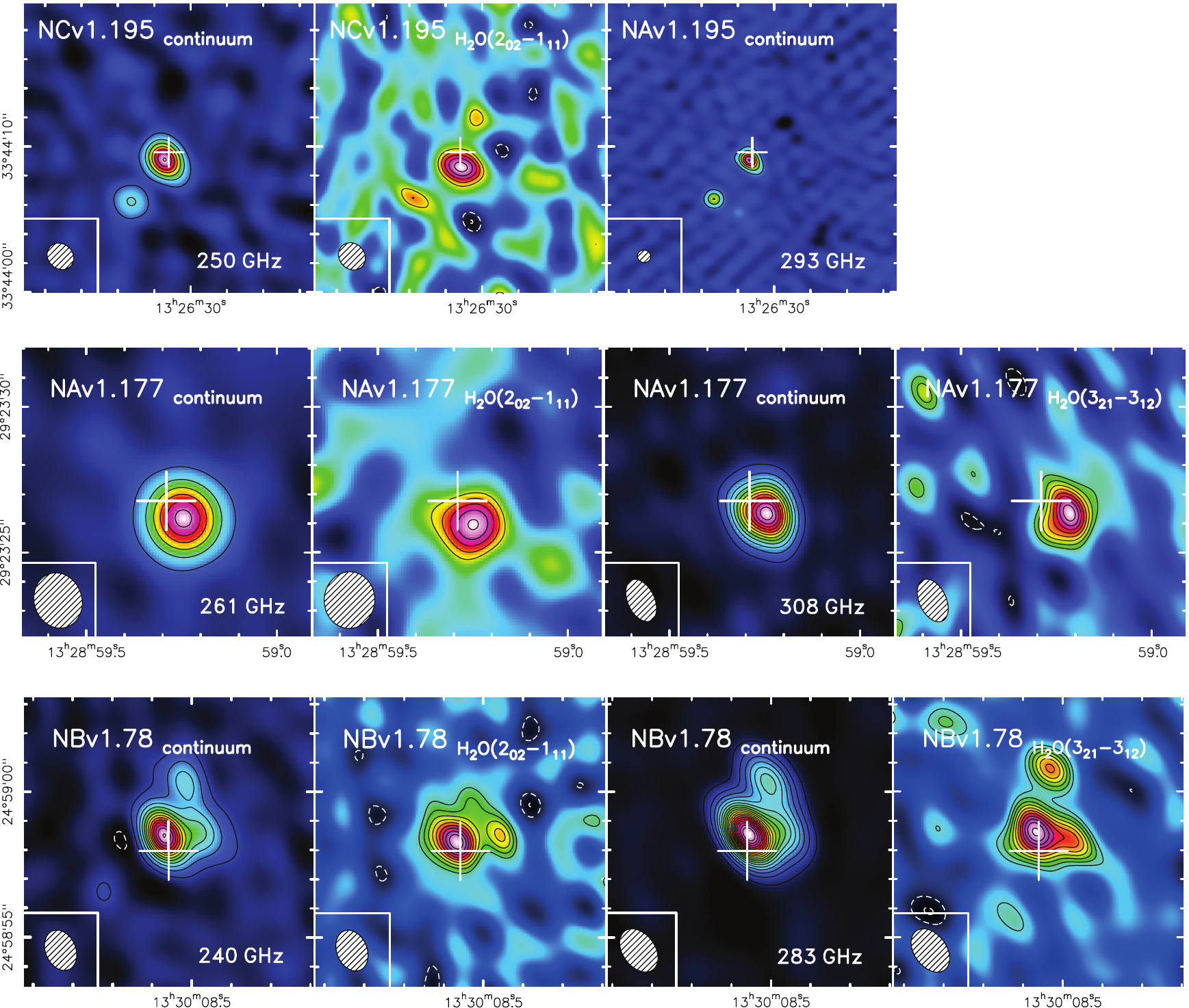}
 \caption{
		  (See Fig.\,\ref{fig:map-1} caption.) For each observation, the $1\,\sigma$ contour for 
		  the continuum (mJy\,beam$^{-1}$) and the \hto\ emission line (Jy\,km\,s$^{-1}$\,beam$^{-1}$) 
		  are as follows: NCv1.195 \htot202111 (0.34/0.51), NCv1.195 \htot321312 
		  (0.48/--), NAv1.177 \htot202111 (0.58/0.65), NAv1.177 \htot321312 (0.38/0.58), 
		  NBv1.78 \htot202111 (0.28/0.30), NBv1.78 \htot321312 (0.21/0.29).
		  }
   \label{fig:map-2}
      \end{center}
\end{figure*}

 \begin{figure*}[htbp]
	 \begin{center}
 \includegraphics[scale=1.0]{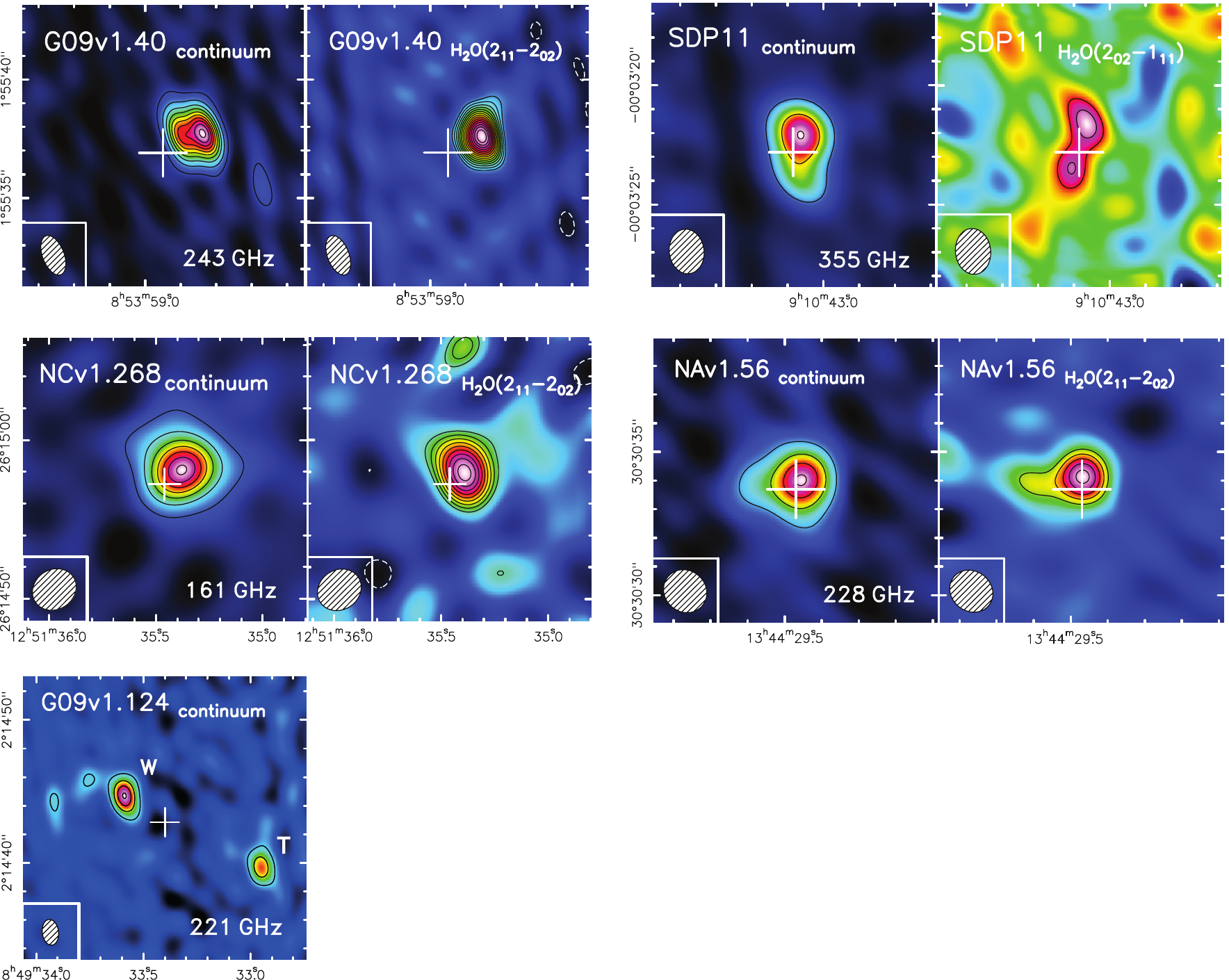}
 \caption{
		  (See Fig.\,\ref{fig:map-1} caption.) For each observation, the $1\,\sigma$ contour for 
		  the continuum (mJy\,beam$^{-1}$) and the \hto\ emission line 
		  (Jy\,km\,s$^{-1}$\,beam$^{-1}$) are as follows: G09v1.124 \htot211202 (0.17/--), 
		  G09v1.40 \htot211202 (0.19/0.32), SDP11 \htot202111 (1.30/1.04), NCv1.268 
		  \htot211202 (0.13/0.39) and NAv1.56 \htot211202 (0.53/1.02).
		  }
  \label{fig:map-3}
     \end{center}	 
 \end{figure*}
 \label{fig:map-all}
\end{subfigures}


\label{Individual sources}
In the appendix, we describe the observational results of 
each source, including the lensing model, the \hto\ spectrum, 
mapping of the \hto\ and continuum emission (by showing 
low-resolution NOEMA \hto\ and continuum images), the ratio 
between velocity-integrated flux densities of different \hto\ 
transitions, and the comparison between \hto\ and CO emission.

\subsection{G09v1.97 at $z=3.634$}
\label{G09v1.97}
The galaxy G09v1.97 has the second largest redshift in our sample 
obtained by CARMA (Riechers et al., in prep.). In the SMA 880\,$\mu$m 
image \citepalias{2013ApJ...779...25B}, similar to SDP81 
\citep{2015ApJ...808L...4A, 2015MNRAS.452.2258D}, it displays 
a triple arc structure with an angular size scale of $\sim2$\,$''$. 
However, there are two foreground deflectors at two different 
redshifts, making this complex mass distribution a very unusual 
case. \citetalias{2013ApJ...779...25B} estimate a lensing 
amplification $\mu=6.9\pm0.6$.

We have observed both para-\htot211202 and ortho-\htot321312 
lines at 162\,GHz and 251\,GHz, respectively. The source is 
clearly unresolved at 162\,GHz, but marginally resolved at 
251\,GHz as displayed in Fig.\,\ref{fig:map-all}. The ratio 
between the peak and the spatially integrated flux density of 
the continuum ($S_{\nu}(\mathrm{ct})^{\mathrm{pk}}/S_{\nu}(\mathrm{ct})$) 
is $0.95\pm0.03$ and $0.60\pm0.01$ at 162\,GHz and 251\,GHz, 
respectively. The \hto\ emission line peak to spatially integrated 
flux density ratio ($S_\mathrm{H_2O}^{\mathrm{pk}}/S_\mathrm{H_2O}$)
for $J=2$ and $J=3$ are $1.0\pm0.2$ and $0.5\pm0.2$, respectively.
Therefore, the spatial concentrations of \hto\ and continuum image
are in good agreement within the uncertainties.

Both the continuum and the \hto\ lines are well detected in 
this source. The two \hto\ lines are well fitted by single 
Gaussian profiles with similar linewidths ($257\pm27$ and 
$234\pm34$\,km\,s$^{-1}$, Fig.\,\ref{fig:spectra-all} 
and Table\,\ref{table:h2o_properties}). The difference in 
linewidth (23\,km\,s$^{-1}$) is smaller than the errors of 
the linewidth. Therefore, there is no significant difference 
between the spectra of the two transitions. The ratio between 
\ihtot321312 and \ihtot211202 is $0.91\pm0.12$, which is the 
lowest of our five detected sources in both lines. However, 
this ratio remains consistent with the observations of local 
galaxies \citepalias[][and Fig.\ 4]{2013ApJ...771L..24Y}, by 
taking the uncertainty into account.

From our CO line observations we find a line FWHM of 
\wco\,=\,$224\pm32$ and $292\pm86$\,km\,s$^{-1}$ for the \co54 
and \co65 lines, respectively, which are within 1\,$\sigma$ to 
the \hto\ FWHMs.
The observed ratio of \ihto/\ico\ for both the \co54 and 
\co65 lines, is about 0.4 with less than 25\% uncertainty.

We have tentatively detected an \htop\ line in this source as well 
(see Section \ref{htop}).

\subsection{G12v2.43 at $z=3.127$}
The source is marginally resolved in the SMA 880\,$\mu$m image 
\citepalias[Figure 2 of][]{2013ApJ...779...25B}, with a size 
$\sim1.5$\,$''$, but there is no obvious strongly lensed structures 
such as multiple images. It is not yet possible to build a lensing 
model for this source because the search for a deflector by 
\citetalias{2013ApJ...779...25B} has been unsuccessful.


Both para-\htot202111 and ortho-\htot321312 lines are well 
detected, as shown in Fig.\,\ref{fig:map-all}, and the source is 
unresolved, consistent with the SMA image. The ratios of 
$S_{\nu}(\mathrm{ct})^{\mathrm{pk}}/S_{\nu}(\mathrm{ct})$ 
for 239\,GHz and 282\,GHz are $0.71\pm0.02$ and $0.87\pm0.01$, 
respectively, while the $S_\mathrm{H_2O}^{\mathrm{pk}}/S_\mathrm{H_2O}$ 
for \htot202111 and \htot321312 are $0.6\pm0.2$ and $1.0\pm0.2$, 
respectively.

The two \hto\ lines are both well fitted by a single Gaussian 
profile. The FWHMs are $201\pm27$ and $221\pm20$\,km\,s$^{-1}$ 
for para-\htot202111 and ortho-\htot321312, respectively.
The difference is within the 1\,$\sigma$ uncertainty. 

The velocity-integrated flux density ratio of high-lying 
over low-lying \hto\ line of this source, 
\ihtot321312/\ihtot202111\;$=1.2\pm0.2$, which is slightly 
lower than that of Arp\,220 as shown in Fig.\,\ref{fig:h2o-sled}. 
The linewidths of the \hto\ lines ($201\pm27$ and $221\pm20$\,km\,s$^{-1}$)
are the narrowest ones among our sources. The values are also 
very close to the \co10 linewidth 
\citep[$210\pm30$\,km\,s$^{-1}$,][]{2012ApJ...752..152H}. 
Their similarity indicates that there is not likely any strong 
differential lensing effect between the CO 
and \hto\ emissions in this case.

\subsection{NCv1.143 at $z=3.565$}
Having a redshift of $z=3.565$ from CO observation by CARMA 
(Riechers et al., in prep.), this source is one of the brightest 
(at submm) in our sample. It is resolved by the SMA 880\,$\mu$m 
beam \citepalias{2013ApJ...779...25B} with a size $\sim2$\,$''$, 
featured by two components at the northeast and southwest 
directions. With a single deflector, the lensing model estimates 
a magnification factor of $\mu=6.9\pm0.6$.

As displayed in Fig.\,\ref{fig:map-all}, both the lines and the 
continuum are very strong and well detected. The ratio 
$S_{\nu}(\mathrm{ct})^{\mathrm{pk}}/S_{\nu}(\mathrm{ct}) = 0.86\pm0.02$
shows that the source is unresolved at 165\,GHz 
(for observing para-\hto\ line \t211202). At 255\,GHz, the ratios 
$S_{\nu}(\mathrm{ct})^{\mathrm{pk}}/S_{\nu}(\mathrm{ct})=0.55\pm0.01$ 
and $S_\mathrm{H_2O}^{\mathrm{pk}}/S_\mathrm{H_2O}=0.7\pm0.2$ indicate 
that the source is partially resolved, consistent with the SMA result. 

Both the \htot202111 and \htot321312 lines can be fitted by single 
Gaussian profiles. The ratio of \ihtot321312/\ihtot211202 is $1.36\pm0.13$, 
close to the mean ratio found in the nearby star-forming-dominated 
galaxies \citepalias[][and Fig.\,\ref{fig:h2o-sled}]{2013ApJ...771L..24Y}. 
The linewidths of \htot211202 and \htot321312 are $293\pm15$ and 
$233\pm22$\,km\,s$^{-1}$, respectively. Although the former is larger, 
they are compatible within an uncertainty of 1.6\,$\sigma$. Also, the 
\hto\ linewidth agrees well with the \co54 and \co65 linewidths 
($273\pm27$ and $284\pm27$\,km\,s$^{-1}$, see 
Table\,\ref{table:co_properties}). Therefore, the line ratios are 
unlikely to be affected by differential magnification. The observed 
ratio of \ihto/\ico\ is 0.4--0.5 and 0.6--0.7 (uncertainties are within 
13\%) for the $J=2$ para-\hto\ and $J=3$ ortho-\hto, respectively.

We have also detected ortho-\htop(\t211202) and ortho-\htop(\t202111) 
fine structure lines together with para-\htot211202 in this source. The 
further discussion of the \htop\ spectra and its interpretation are 
given in Section \ref{htop}.

\subsection{NAv1.195 at $z=2.951$}
As quoted in \citetalias{2013ApJ...779...25B}, the redshift of 
this source was first obtained by the CO observation (Harris et 
al., in prep.). Its SMA image shows a typical lensed feature with 
two components separated by $\sim2$\,$''$ along the northwest 
and southeast direction. The lensing model suggests a modest 
magnification factor $\mu=4.1\pm0.3$.

We have robust detections of \htot202111 and the continuum emission 
at 250\,GHz and 293\,GHz (Fig.\,\ref{fig:map-all}). However, the 
\htot321312 line is not detected (Fig.\,\ref{fig:spectra-all}), 
at odds with the other five sources. 
Therefore, we only show the image of the dust continuum emission at 
this frequency in Fig.\,\ref{fig:map-all}. The source is clearly 
resolved into two components in the three images, and the northwest 
component is about 4 times stronger than the southeast one in the 
continuum images, in agreement with the SMA image 
\citepalias{2013ApJ...779...25B}. For the continuum at 250\,GHz, 
the peak to total integrated flux density ratio  
$S_{\nu}(\mathrm{ct})^{\mathrm{pk}}/S_{\nu}(\mathrm{ct})=0.54\pm0.02$, 
and for \htot202111, the ratio  
$S_\mathrm{H_2O}^{\mathrm{pk}}/S_\mathrm{H_2O}$ equals to $0.6\pm0.3$. 
Therefore, the spatial distributions of dust and the \hto\ emission are 
likely to be similar in this source. In the observation at 293\,GHz, 
$S_{\nu}(\mathrm{ct})^{\mathrm{pk}}/S_{\nu}(\mathrm{ct})=0.42\pm0.01$, 
due to a smaller synthesis beam (Table\,\ref{table:obs_log}). 

Fig.\,\ref{fig:spectra-all} shows the spectra corresponding to 
the two observations of NAv1.195. The \htot202111 line can be 
fitted by a single Gaussian profile, with a linewidth equal to 
$328\pm51$\,km\,s$^{-1}$. We have not detected the higher 
excitation \htot321312 line as mentioned above. By assuming the 
same linewidth as the lower$-J$ \hto\ line, we can infer a 
2\,$\sigma$ detection limit of 2.56\,Jy\,km\,s$^{-1}$. 
This yields a ratio of \htot321312/\htot202111 $\lesssim0.6$. 
This value is significantly lower than that in the five other 
sources where it ranges from 0.75 to 1.60 (errors are within 
25\%), but it remains close to the lowest values measured in 
local galaxies \citepalias{2013ApJ...771L..24Y} as shown in 
Table \ref{table:Lir_Lh2o_ratios} and Fig. \ref{fig:h2o-sled}. 
This low ratio of \hto\ lines probably originates 
from different excitation conditions, especially for the \fir\ 
radiation field, since the line \htot321312 is mainly populated 
through \fir\ pumping via absorbing 75\,$\mu$m photons (see 
Section \ref{fig:h2o-model}). The \co54 line of the source has a 
linewidth of $281\pm16$\,km\,s$^{-1}$, which is comparable with the 
\hto\ line profile. The observed ratio of \ihto/\ico\ (\co54) is $\leq0.4$.

\subsection{NAv1.177 at $z=2.778$}
NOEMA observation of the CO line in this source gives a redshift of 
$z$\,=\,2.778 (Krips et al., in prep.). The SMA 880\,$\mu$m image 
shows a compact structure with two peaks $\sim1$\,$''$ away along 
the eastwest direction, and the western component is the dominant 
one \citepalias[Figure 2 of][]{2013ApJ...779...25B}. However, due 
to the absence of deflector in the foreground optical image 
from SDSS and lack of the deep optical and near-\ir\ images, 
the lensing properties are still unknown \citepalias{2013ApJ...779...25B}.

As displayed in Fig.\,\ref{fig:map-all}, both the lines of 
\htot202111 and \htot321312 and the corresponding dust 
continuum are well detected. The ratios 
$S_{\nu}(\mathrm{ct})^{\mathrm{pk}}/S_{\nu}(\mathrm{ct})$ are 
$0.75\pm0.02$ and $0.62\pm0.01$ for observation at 261\,GHz and 
308\,GHz, respectively. One should notice that the direction of 
the synthesised beam is perpendicular to the alignment of the two 
components in the image, thus the source is marginally resolved in the 
\htot202111 and the corresponding dust continuum images. For 
the \htot321312 observation at higher frequency, we can see 
the partially resolved feature. The peak to total flux ratios of 
\hto\ are $S_\mathrm{H_2O}^{\mathrm{pk}}/S_\mathrm{H_2O}\sim 0.8\pm0.3$ 
and $0.6\pm0.1$ for the the \htot202111 and \htot321312 
lines, respectively, indicating similar spatial distribution 
compared with the dust emission. 

The \hto\ spectra displayed in Fig.\,\ref{fig:spectra-all} show 
single Gaussian profiles with $FWHM\,= 241\pm41$ and 
$272\pm24$\,km\,s$^{-1}$ (Table\,\ref{table:h2o_properties}).  
The profiles of the two \hto\ lines are similar within the 
uncertainties. The line ratio, \ihtot321312/\ihtot202111\,$=1.34\pm0.24$. 
This value is close to that found in Arp\,220 and it is the 
largest ratio in our sample. We have also detected the \co32 
and \co54 lines using the IRAM 30m telescope in this source 
(Table\,\ref{table:co_properties}), the linewidth of CO 
is around $230\pm16$\,km\,s$^{-1}$ which is similar to 
the width of the detected \hto\ lines. The ratio 
of \ihto/\ico\ is from 0.5 to 1.1 with less than 20\% uncertainties.

\subsection{NBv1.78 at $z=3.111$}
The CO redshift of NBv1.78 was obtained by Riechers et al. (in prep.) 
via CARMA, $z$\,=\,3.111, and the data of the \htot202111 line were 
obtained by \citetalias{2013A&A...551A.115O}. The source is resolved 
in the SMA 880\,$\mu$m image \citepalias{2013ApJ...779...25B} with 
a somewhat complex morphology, and the size is $\sim2.5$\,$''$. There 
are three main components in the image. The two strong components located at 
northwest and southeast direction of the image, and the weakest 
component close to the southeast. The derived lensing magnification 
is $\mu = 13.5\pm1.5$, which is the second largest among our 
sample. In the near-\ir\ images, the source has a similar three-component 
Einstein ring-like structure with a slightly smaller magnification 
$\mu = 10.8^{+0.3}_{-0.2}$ \citep{2014ApJ...797..138C}.
 
Our NOEMA images of both continuums and \hto\ lines as shown in 
Fig.\,\ref{fig:map-all} are consistent with the SMA 880\,$\mu$m 
image. The images are resolved into two main parts, while the 
southern component is extended along the western side. The continuum 
and \hto\ line images have fairly high S/N. From the observation 
of \htot202111 at 241\,GHz (\citetalias{2013A&A...551A.115O}, 
note that this observation was performed at higher resolution 
with a 1.4\,$''$\,$\times$\,1.0\,$''$ beam), the values of 
$S_{\nu}(\mathrm{ct})^{\mathrm{pk}}/S_{\nu}(\mathrm{ct})$ 
and $S_\mathrm{H_2O}^{\mathrm{pk}}/S_\mathrm{H_2O}$ agree 
well, which are $0.42\pm0.01$ and $0.4\pm0.1$, respectively. 
The continuum image at 283\,GHz gives 
$S_{\nu}(\mathrm{ct})^{\mathrm{pk}}/S_{\nu}(\mathrm{ct})=0.69\pm0.01$, 
and the image of \htot321312 gives 
$S_\mathrm{H_2O}^{\mathrm{pk}}/S_\mathrm{H_2O}=0.8\pm0.1$. 
The similarity of the peak to spatially integrated flux 
density ratios suggest that the spatial distribution of \hto\ 
and submm dust continuum are likely to be consistent. 
Additionally, the images of $J=2$ and $J=3$ images are 
also consistent within the uncertainty.

NBv1.78 has a very broad linewidth compared with the other 
sources. As shown in Fig.\,\ref{fig:spectra-all}, the linewidth 
of \htot202111 and \htot321312 are $510\pm90$ and $607\pm43$\,km\,s$^{-1}$,
respectively. The two lines have similar profiles. The source has a
\ihtot321312/\ihtot202111 ratio equal to 1, within the 
range of the local galaxies (Fig.\,\ref{fig:h2o-sled}). The \co54 
and \co65 observations (Table\,\ref{table:co_properties}) give 
linewidths of $614\pm53$ and $734\pm85$\,km\,s$^{-1}$, which are 
wider than the \hto\ lines. The observed ratio of \ihto/\ico\ 
is about 0.7 with less than 25\% uncertainty for $J=2$ \hto.

\subsection{G09v1.40 at $z=2.093$}
A CO redshift of G09v1.40, $z=2.0894$ was obtained by CSO/Z-Spec 
(Lupu et al., in prep.). But, through our \hto\ observation, we 
find a redshift of $z=2.093$. Our value is consistent with the 
\co32 observation by Riechers et al. (in prep), and we have 
adopted this value. SMA observation of the 880\,$\mu$m dust 
continuum shows two close components with a separation of 
$\sim 1$\,$''$ along the east and west direction. The lensing 
model estimates $\mu=15.3\pm3.5$, which is the largest 
magnification in our sample. The Keck near-\ir\ image of the 
source suggests a magnification of $\mu=11.4^{+0.9}_{-1.0}$ 
for the stellar component \citep{2014ApJ...797..138C}.

The \htot211202 line is well detected as well as the corresponding 
dust continuum. As shown by the images of the \hto\ and dust 
continuum (Fig.\,\ref{fig:map-all}), the source is partially 
resolved by the synthesised beam. The two component (east and west) 
structure is consistent with the 880\,$\mu$m image, and the western 
component is stronger than the eastern one. Both ratios of 
$S_{\nu}(\mathrm{ct})^{\mathrm{pk}}/S_{\nu}(\mathrm{ct})$ 
and $S_\mathrm{H_2O}^{\mathrm{pk}}/S_\mathrm{H_2O}$ are found to 
be 0.6 (uncertainty $<13\%$). However, the eastern component is 
not detected in the \hto\ image.

The \htot211202 line can be fitted with a single Gaussian profile 
with a FWHM of $277\pm14$\,km\,s$^{-1}$ (Fig.\,\ref{fig:spectra-all}). 
However, the line has a steeper decline on the red side of the spectrum. 
The \co43 observation gives a linewidth of $198\pm51$\,km\,s$^{-1}$, 
which is $0.7\pm0.2$ times narrower than that of the \hto\ line. The 
velocity-integrated flux density of the \hto\ is larger than that of 
the \co43 in this source with a ratio of \ihto/\ico\,$=1.1\pm0.3$.

\subsection{SDP11 at $z=1.786$}
The CO observation by \cite{2012ApJ...757..135L} found $z$\,=\,1.786. 
The SMA 880\,$\mu$m image displays a typical strongly lensed morphology 
with two components, north and south, respectively. The size of the 
source is $\sim2$\,$''$. The gravitational magnification estimated 
by \citetalias{2013ApJ...779...25B} is $\mu = 10.9\pm1.3$.

As shown in Fig.\,\ref{fig:map-all}, the source is partially resolved. 
The dust continuum image shows an extended structure along the north 
and south direction, with the brightest peak in the northern part. 
The noisy images are consistent with the SMA 880\,$\mu$m observation. 
The ratio of $S_{\nu}(\mathrm{ct})^{\mathrm{pk}}/S_{\nu}(\mathrm{ct})$ 
is $0.56\pm0.03$. The image shows two distinctive components 
in the north and south direction. This structure also agrees 
with the high resolution SMA 880\,$\mu$m image. Additionally, 
$S_\mathrm{H_2O}^{\mathrm{pk}}/S_\mathrm{H_2O}=0.4\pm0.2$ suggests that 
the \hto\ image is slightly resolved compared with the dust emission. 
This difference may come from their different spatial distribution.

Although the noise level of its spectrum is the highest among our sources
because of the high frequency, namely 355\,GHz, the \htot202111 line 
is marginally detected with $S/N$\,=\,4.6. The linewidth is 
$214\pm41$\,km\,s$^{-1}$ (Fig.\,\ref{fig:spectra-all}).

\subsection{NCv1.268 at $z=3.675$}
The redshift of NCv1.268 was obtained by the CO observation of 
Krips et al. (in prep.). A typical strongly lensed morphology was 
found by the SMA 880\,$\mu$m observation \citepalias{2013ApJ...779...25B}, 
with a strong arc-like component in the south direction. The structure has 
a size $\sim2.5$\,$''$. The \citetalias{2013ApJ...779...25B} lensing 
model estimates $\mu=13.0\pm1.5$.

The source is marginally resolved by the NOEMA synthesis beam 
(Fig.\,\ref{fig:map-all}). When comparing the flux ratios 
between the dust and \hto\ emission from the peak and the 
spatially integrated values, they give 
$S_{\nu}(\mathrm{ct})^{\mathrm{pk}}/S_{\nu}(\mathrm{ct})=0.66\pm0.01$ 
and $S_\mathrm{H_2O}^{\mathrm{pk}}/S_\mathrm{H_2O}=0.6\pm0.1$. 
The values of dust emission and the \hto\ image are in good agreement. 

NCv1.268 is the only source in which we have detected a double-peaked 
line profile from our new observations, with a slightly stronger blue 
velocity component (Fig.\,\ref{fig:spectra-all}). The total linewidth 
is very large, \,$731\pm75$\,km\,s$^{-1}$.


\subsection{NAv1.56 at $z=2.301$}
\cite{2012ApJ...752..152H} give a CO redshift of $z=2.3010$ for 
this source. The SMA 880\,$\mu$m dust continuum image shows a 
classic strongly lensed morphology with multiple images. It 
consists of an arc-like component in the western direction 
and a point-like component toward the east. They are separated 
$\sim2$\,$''$. The lens model implies that the magnification 
factor $\mu=11.7\pm0.9$ \citepalias{2013ApJ...779...25B}.

As shown in Fig.\,\ref{fig:map-all}, the source is marginally 
resolved, with an extended morphology in the eastern part. 
The structures displayed in the dust and \hto\ images are 
similar. The ratios 
$S_{\nu}(\mathrm{ct})^{\mathrm{pk}}/S_{\nu}(\mathrm{ct})=0.62\pm0.03$ 
and $S_\mathrm{H_2O}^{\mathrm{pk}}/S_\mathrm{H_2O}=0.7\pm0.2$ 
also suggest their similarity within the errors. Most of the 
fluxes are concentrated in the western part, which agrees 
with the SMA 880\,$\mu$m image.

The \htot211202 line can be fitted by a single Gaussian with a large 
linewidth equal to $593\pm56$\,km\,s$^{-1}$. The \co43 line observation by 
NOEMA (Oteo et al., in prep.) gives a linewidth of $621\pm47$\,km\,s$^{-1}$. 
The linewidths of CO and \hto\ are in very good agreement. Our noisy detection 
of \co54 at IRAM 30m (Table\,\ref{table:co_properties}) gives a ratio of
\ihto/\ico\;$=0.8\pm0.3$.

\subsection{G09v1.124 at $z=2.410$}\label{g09v1.124}
The redshift of this source is measured by CO observation 
\citep{2012ApJ...752..152H}. This multiple source, with two main 
components, each with intrinsic \lir\;$> 10^{13}$\,\lsun, separated 
by 10" (Fig.\,\ref{fig:map-all}), was studied in detail by 
\citep[][see also \citetalias{2013ApJ...779...25B} and Oteo et al., 
in prep.]{2013ApJ...772..137I}. It is special in our sample since 
the two main sources are from two very different HyLIRGs rather than 
multiple images of a single source generated by strong gravitational 
lensing. The eastern component G09v1.124-W, which contains a powerful 
AGN (Oteo et al., in prep.), is unlensed and the western component 
G09v1.124-T is only weakly lensed with a magnification factor 
$\mu = 1.5 \pm 0.2$. Thus, throughout the discussions, we treat 
G09v1.124-W and G09v1.124-T as two distinct sources 
(see Tables\,\ref{table:previous_obs_properties}, 
\ref{table:h2o_properties}, \ref{table:Lir_Lh2o} and 
\ref{table:Lir_Lh2o_ratios}).

Probably because of this too small lensing magnification and
the smaller values of each $\mu$\lir\ 
(Table\,\ref{table:previous_obs_properties}), we have only 
detected the dust continuum emission in this source. The 
\htot211202 line is undetected. The 2\,$\sigma$ upper limits
of the velocity-integrated flux density of the \htot211202 
line show that the values of \ihto\ are more than three times 
smaller than in the other sources. As seen in 
Table\,\ref{table:h2o_properties} and Fig.\,\ref{fig:h2o-ir}, 
the ratio \lhto/\lir\ is smaller than all our other sources 
for G09v1.124-W, probably because of its 
strong AGN. However, for G09v1.124-T this ratio, albeit small, 
might be comparable to that of G09v1.97. The dust continuum at 
221\,GHz follows the same structure as the previously published 
observations (Fig.\,\ref{fig:map-all} and \citealt{2013ApJ...772..137I}). 
Both the eastern component (W) and the western one (T) are 
marginally resolved by the synthesised beam. The peak to 
total continuum flux ratios are  
$S_{\nu}(\mathrm{ct})^{\mathrm{pk}}/S_{\nu}(\mathrm{ct})=0.84\pm0.03$ 
and $0.83\pm0.04$, respectively.

\section{\htop\ detections in local ULIRGs} 

The study using the {\it Herschel} SPIRE/FTS spectra of 167 local 
galaxies has revealed several emission and absorption lines 
of \htop, which are ortho-\htop\, lines \t211202\,$_{(5/2-5/2)}$, 
\t202111\,$_{(5/2-3/2)}$, \t111000\,$_{(3/2-1/2)}$, 
\t111000\,$_{(1/2-1/2)}$ 
\citepalias[][see also Table\,\ref{table:htop}]{2013ApJ...771L..24Y}. 
All $J \geq 2$ \htop\ lines are in emission. Table\,\ref{table:htop} 
gives values of the \htop\ flux and luminosity for those among the 
\citetalias{2013ApJ...771L..24Y} sample where \htop\ lines are 
(tentatively) detected with $S/N$\;$\gtrsim 2.5$. However, for the 
\htop(\t111000) lines which connect the ground state, they are often 
found to be in emission in AGN-dominated sources while they are in 
absorption in star-forming-dominated ones. A full description of 
the dataset for this {\it Herschel} SPIRE/FTS survey will be 
given in Liu et al. in prep. At \hz\, prior to our study, the 
$J=2$ ortho-\htop\ doublet lines seem to have only been tentatively 
detected in two sources, SPT0346-52 \citep{2013ApJ...767...88W} 
and HFLS3 \citep{2013Natur.496..329R}.

\begin{table}[htbp]
\setlength{\tabcolsep}{0.28em}
\small
\caption{Beam-matched \htop, \hto\ line and \ir\ luminosities from local detections ({\it Herschel} SPIRE/FTS archive) and \hz\ {\it Herschel} lensed galaxies.}
\centering
\begin{tabular}{lccrr}
\hline
\hline
Source           &  \t202111\,$_{(5/2-3/2)}$ & \t211202\,$_{(5/2-5/2)}$ &  $L_\mathrm{IR-beam}$ &  \lhto           \\
                 &  $10^6$\,\lsun            &  $10^6$\,\lsun           &  $10^{11}$\,\lsun     &  $10^6$\,\lsun   \\
\hline  
\multicolumn{5}{c}{local ULIRGs} \\         
\hline                                           
ESO320-G030      &  $0.4\pm0.1$              &    --                    &   1.5                 &  0.9  \\
CGCG049-057      &  $0.3\pm0.1$              &    $0.7\pm0.2$           &   1.9                 &  1.6  \\
NGC2623          &  $0.5\pm0.2$              &    $0.7\pm0.3$           &   3.5                 &  2.0  \\
Arp299-A         &  $0.8\pm0.2$              &    $0.7\pm0.3$           &   5.4                 &  1.3  \\
Arp220           &  $3.7\pm0.8$              &    $3.5\pm0.9$           &  16.2                 &  15.2  \\
IRAS13120-5453   &  $3.6\pm1.2$              &    $2.4\pm1.0$           &  28.2                 &  10.7  \\
IRASF17207-0014  &  $6.8\pm2.2$              &    $6.3\pm2.2$           &  24.6                 &  15.5  \\
Mrk231           &  $2.6\pm1.0$              &    $3.0\pm1.5$           &  32.4                 &   9.3  \\
\hline                                                                                                
\multicolumn{5}{c}{\hz\ lensed galaxies} \\      
\hline                                                                                         
NCv1.143         &  $28.7\pm7.2$             &    $28.6\pm7.2$          & 113.6                 &  89.7  \\
G09v1.97         &  $35.8\pm9.0$             &     --                   & 211.4                 & 106.6  \\
G15v2.779        &   --                      &    $55.8\pm15.9$         & 270.0                 & 266.0  \\
HFLS3            &   --                      &  $261.8\pm200.3$         & 364.0                 & 737.8   \\
\hline
\end{tabular}
\tablefoot{
        Luminosity of \htop\ fine structure lines 
        \t202111\,$_{(5/2-3/2)}$ (742.1\,GHz), 
        \t211202\,$_{(5/2-5/2)}$ (746.5\,GHz),
		and \htot211202. As for the local sources 
		the {\it Herschel} SPIRE/FTS beam is smaller 
		than the entire source for some galaxies, 
		here $L_\mathrm{IR-beam}$ represents the 
		$L_\mathrm{IR}$ from the same beam area 
		as the measured emission line (a full 
		description will be given by Liu et al. 
		in prep.). \lhto\ is the luminosity of 
		\htot211202. HFLS3 data are taken from \citet{2013Natur.496..329R}.
       }
\label{table:htop}
\end{table}
\normalsize

\end{appendix}

\end{CJK*}
\end{document}